\documentclass[aps,pre,twocolumn,10pt]{revtex4-2}
\usepackage[colorlinks = True, linkcolor = blue, citecolor = blue, breaklinks=True]{hyperref}
\usepackage{times}
 \usepackage{graphicx}
\usepackage{placeins}  
\usepackage{setspace}
\usepackage{calc}
\usepackage{calrsfs}
\usepackage{float}
\usepackage{mathtools}
\usepackage{ulem}
\usepackage{amssymb,amsmath,amsfonts}
\usepackage{graphics,epsfig}
\usepackage{color,bm,hyperref}
\usepackage{ulem} 
\usepackage{xcolor}
\usepackage{tikz}
\usepackage{yfonts}
\usepackage{comment}
\usetikzlibrary{shapes.geometric,arrows}
\usetikzlibrary{decorations.markings}
\DeclareMathAlphabet{\pazocal}{OMS}{zplm}{m}{n}

\begin{abstract}
 We explore the impact of global resetting on Kuramoto-type models of coupled limit-cycle oscillators with distributed frequencies both in absence and presence of noise. The dynamics comprises repeated interruption of the bare dynamics at random times with simultaneous resetting of phases of all the oscillators to a predefined state. To characterize the stationary-state behavior, we develop an analytical framework that spans across different generalizations of the Kuramoto model involving either quenched or annealed disorder or both, and for any choice of the natural frequency distribution. The framework applies to the dynamics both in absence and presence of resetting, and is employed to obtain in particular the stationary-state synchronization order parameter of the system, which is a measure of spontaneous ordering among the oscillator phases. A key finding is the pivotal role of
correlations in shaping the ordering dynamics under resetting.
\end{abstract}
\begin{document}
\title{Stationary-state dynamics of interacting phase oscillators in presence of noise and stochastic resetting}
\author{Anish Acharya$^1$}
\email{Corresponding author; email: anish.acharya@tifr.res.in}
\author{Mrinal Sarkar$^2$}
\email{Email: sarkar@thphys.uni-heidelberg.de}
\author{Shamik Gupta$^1$}
\affiliation{$^1$Department of Theoretical Physics, Tata Institute of Fundamental Research, Homi Bhabha Road, Mumbai 400005, India}
\affiliation{$^2$Institut f\"{u}r Theoretische Physik, Universit\"{a}t Heidelberg, Philosophenweg 19, 69120 Heidelberg, Germany}
\maketitle

\section{Introduction}
\label{sec:intro}
Over the past decade, stochastic resetting has emerged as a powerful mechanism to control and optimize dynamical processes in nonequilibrium set-ups. Repeated interruption and resetting of a dynamics to a given state can alter significantly the resulting properties, both static and dynamic, yielding non-trivial nonequilibrium stationary states and enhancing efficiencies in search processes. Beginning with the seminal paper by Evans and Majumdar~\cite{evans2011diffusion}, stochastic resetting has been extensively studied in a wide variety of scenarios, from classical~\cite{pal2016diffusion,PhysRevE.99.012121,PhysRevE.92.062148,boyer2019anderson,PhysRevE.101.062147,ray2021mitigating,evans2022exactly,gupta2014fluctuating,basu2019symmetric,karthika2020totally} to quantum~\cite{mukherjee2018quantum,das2022quantum,acharya2023tight,yin2023restart,perfetto2021designing}, and from chemical~\cite{PhysRevE.92.060101,PhysRevResearch.3.013273} to biological~\cite{roldan2016stochastic,tucci2020controlling,reuveni2016optimal,boyer2014random,giuggioli2019comparison,boyer2014random}; for an overview, we refer the reader to recent reviews~\cite{evans2020stochastic,gupta2022stochastic,nagar2023stochastic}.

Even at the level of dynamics of a single particle, stochastic resetting introduces many non-trivial features. One example is a single Brownian particle subject to resetting, wherein the Brownian particle, while undergoing its intrinsic stochastic dynamics from a given initial condition, is repeatedly reset to the initial condition at random times, after which the dynamics starts afresh. In the absence of resetting, the position distribution of the Brownian particle spreads indefinitely in time, and the dynamics does not ever attain any stationary state. However, resetting effectively confines the particle within a finite region, resulting in a non-trivial nonequilibrium stationary state (NESS)~\cite{evans2011diffusion}. 

When extended to a system of many independent (noninteracting) Brownian particles, simultaneous resetting leads to strong correlations; the dynamics serves as one of the very few solvable set-ups to understand the role of strong correlations in an NESS~\cite{biroli2023extreme}. The complexity of the dynamics and hence of correlations is further enhanced when the particles are taken to be interacting. Addressing the interplay of many-body effects with resetting dynamics has prompted growing interests in study of resetting in many-body interacting systems~\cite{nagar2023stochastic}. In this backdrop, we focus here on the Kuramoto model of coupled nonlinear oscillators as a representative many-body system.

A paradigmatic model in its own right, the Kuramoto model~\cite{kuramoto1984chemical} has been widely employed over the years in studying spontaneous synchronization in systems of interacting nonlinear oscillators~\cite {strogatz2000kuramoto,acebron2005kuramoto,pikovsky2015dynamics}. The model comprises a population of limit-cycle oscillators with distributed natural frequencies, coupled all-to-all through a sinusoidal interaction that depends on the difference of phases between the oscillators. Suitable variations of the basic framework of the Kuramoto model have clarified the conditions for occurrence of synchronization in diverse set-ups, including fireflies, cardiac pacemaker cells, and Josephson junction arrays, as also in discussing animal flocking behavior, pedestrian movement on footbridges, rhythmic applause in concert halls, and electrical power distribution networks~\cite{pikovsky2001synchronization,strogatz2004sync}. The main theoretical breakthrough achieved by Kuramoto was the analytical demonstration of the fact that the dynamics at long times allows for the existence of a synchronized phase, namely, a phase in which the oscillator phases evolve in time while maintaining phase differences that are constant in time. Such a phase emerges spontaneously and for any reasonable distribution of oscillator frequency distribution, provided the strength of coupling between the oscillators exceeds a critical value. 

On the theoretical front, the Kuramoto model poses a challenge for a detailed analytical study, particularly when resetting is introduced. In the current context, resetting involves interrupting the bare dynamics at random time intervals $\tau$ distributed according to an exponential, $p(\tau)=\lambda e^{-\lambda \tau}$, at which the phases of all the oscillators are simultaneously reset to a common value. Thus, the resetting move, which is evidently global in nature, takes the system instantaneously to a completely synchronized state. Here, $\lambda >0$ is the resetting rate. Besides the quenched disorder arising from distributed natural frequencies of the oscillators and the nonlinear nature of interaction between the oscillators, further non-trivialities arise from correlations induced due to simultaneous resetting of all the oscillator phases to a common value. This area of research is still in its infancy, with only a few studies reported in recent times~\cite{sarkar2022synchronization,bressloff2024global,bressloff2024kuramoto}.

Our analysis furthers the aforesaid body of work, in that we bring in an additional element, namely, that of annealed disorder in the dynamics, with an attempt to unveil the interplay of three distinct sources of disorder: quenched disorder arising from the distributed natural frequencies in the bare dynamics, temporal disorder introduced through global resetting of the oscillator phases at random times, and annealed disorder introduced through inclusion of Gaussian, white noise in the bare dynamics. The models we study, which are all generalizations of the basic Kuramoto model, are (i) the Brownian mean-field (BMF) model~\cite{chavanis2014brownian,gupta2018statistical}, in which only annealed disorder is present, (ii) the Kuramoto model, in which only quenched disorder is present, and (iii) the noisy Kuramoto model~\cite{sakaguchi1988cooperative}, in which both annealed and quenched disorder are present. We study within a common framework the long-time fate of all the three models both in absence and presence of resetting. 

The Kuramoto model in presence of resetting has been studied in recent past. For example, in Ref.~\cite{sarkar2022synchronization}, it was unveiled how resetting dramatically modifies the phase diagram of the model, allowing for the emergence of a synchronized phase even in parameter regimes for which the bare model does not support such a phase. The theoretical analysis, though exact, was restrictive in its applicability. Namely, in the analysis, the natural frequency distribution characterizing the source of quenched disorder was taken to be Lorentzian. Moreover, in the limit of a very large number of oscillators in which the analysis was done, the single-oscillator distribution, the main element of the analysis, was taken to reside on a particular manifold in the space of all possible single-oscillator distribution functions. The manifold, known as the Ott-Antonsen (OA) manifold, is such that the coefficients in the Fourier expansion of the distribution function bear a special relationship with one another. These relationships constitute the basis of the so-called OA-ansatz~\cite{ott2008low}. For the bare model, it is known that dynamics initiated on this manifold continues to remain on the manifold as time progresses, allowing for an effective description of the synchronization order parameter of the system in terms of a single ordinary differential equation that admits a closed-form solution. 

On introducing resetting, the theoretical analysis pursued in Ref.~\cite{sarkar2022synchronization} assumed that the aforementioned scenario continues to be true: the initial condition was chosen to lie on the OA manifold, and the resetting protocol was chosen to be such as to confine the dynamics always to the manifold as time progresses. This assumption, restrictive as it is, allowed for an exact analytical characterization of the stationary state of the dynamics. Besides extending the resetting protocol to apply to more generalized versions of the Kuramoto model, as mentioned above, a motivation of the current work is to propose a theoretical framework that goes beyond the OA-manifold-based analysis and allows to provide analytical expressions for the synchronization order parameter in the stationary state even when the frequency distribution is not Lorentzian and the dynamics does not necessarily remain confined to the OA manifold.

The analytical framework we propose is based on a combination of a mean-field approach and an approach to solve tridiagonal vector recurrence relations developed in Ref.~\cite{risken1980solutions}. The method allows to compute the stationary-state order parameter with and without stochastic resetting. Our results on the stationary order parameter in absence of resetting reproduce (i) known closed-form expressions for the BMF and the Kuramoto models, see Eq.~\eqref{eq:BMF-rss-consistent} and Fig.~\ref{fig:BMF_globalresettting} for the former and Eqs.~\eqref{eq:rnonzero-app} and~\eqref{Eq:rss_uniform} and Fig.~\ref{fig:Kuramoto_globalresettting} for the latter; (ii) known results for the noisy Kuramoto model, Fig.~\ref{fig:NK_globalresettting}. In this regard, it is noteworthy that our results for the Kuramoto model are obtained without invoking the OA-ansatz, and thus go beyond the choice of a Lorentzian natural frequency distribution for which the ansatz was proposed and studied. 

Next, our results on the stationary-state order parameter in presence of resetting are shown in Figs.~\ref{fig:BMF_globalresettting},~\ref{fig:Kuramoto_globalresettting},~\ref{fig:NK_globalresettting}  for the BMF, the Kuramoto, and the noisy Kuramoto model, respectively. As may be observed, in presence of resetting, one has a qualitative match between our analytical results and results from direct numerical simulations. We attribute this feature to correlations induced in the dynamics by the process of resetting, which is most pronounced at intermediate values of the resetting rate $\lambda$.  The main consequence of resetting, observed across the three different models, is that even in a parameter regime where the corresponding bare dynamics does not allow for a synchronized phase in the stationary state, introducing resetting to a synchronized state does allow for the existence of such a phase. 

The paper is organized as follows. Section~\ref{sec:single-particle} discusses the dynamics of a single particle diffusing on a circle in the absence and presence of stochastic resetting. Kuramoto-type models can also be interpreted as describing the dynamics of particles moving on a unit circle and interacting via a mean-field potential, with the system driven out of equilibrium due to application of external torques that are quenched-disordered random variables. The main reason for including Section~\ref{sec:single-particle} is to demonstrate that nontrivial NESS may be generated for motion on the circle even when the particles are non-interacting. Besides, results from this section help in clarifying analytically the validity of the mean-field approximation that we invoke in the following sections to study our models of interest. Section~\ref{sec:many-particles} extends our analysis to a general interacting-system set-up that encompasses the three models of our interest, namely, the BMF, the Kuramoto, and the noisy Kuramoto model; in this section, we describe our analytical method to characterize the stationary-state behavior in the general set-up. Analysis of the BMF model is pursued in Sec.~\ref{sec:BMF}, that of the Kuramoto model in Sec.~\ref{sec:Kuramoto}, and that of the noisy Kuramoto model in Sec.~\ref{sec:Noisy_Kuramoto}. We assess the validity of our approach and in particular discuss about resetting-induced correlations in Sec.~\ref{sec:MF-validation}, while Sec.~\ref{sec:Conclusion} concludes our work with a discussion of potential future directions. Details of calculations are relegated to Appendices~\ref{sec1:app1} and~\ref{sec2:app1}.

\section{Resetting of a single particle diffusing on a circle}
\label{sec:single-particle}

\subsection{Dynamics in absence of resetting}
\label{sec:single-particle-no-reset}

The equation describing the diffusion of a single particle on a unit circle is given by
\begin{align}
    \frac{d\theta}{dt}=\sqrt{2D}~\eta(t);~D>0,
    \label{eq:bare-dynamics}
\end{align}
with $\theta \in [-\pi,\pi]$ being the angular position of the particle on the circle, and $\eta(t)$ being Gaussian, white noise with zero mean and correlations given by $\langle \eta(t)\eta(t')\rangle=\delta(t-t')$. Here and in the rest of the paper, angular brackets denote averaging over noise realizations. The constant $D$ in Eq.~\eqref{eq:bare-dynamics} sets the strength of the noise effect in the dynamics. 

Let $P_0(\theta,t)$ denote the probability density for the particle to be found between angles $\theta$ and $\theta+d \theta$ on the circle at time $t$. The probability density is normalized as $\int_{-\pi}^{\pi}d\theta~P_0(\theta,t)=1~\forall~t$, is $2\pi$-periodic in $\theta$, and which evolves in time according to the Fokker-Planck equation
\begin{align}
    \frac{\partial P_0(\theta,t)} {\partial t}=D \frac{\partial^2 P_0(\theta,t)} {\partial \theta^2}\label{eq:FK_diff_ring}.
\end{align}
 Performing the Fourier expansion with respect to $\theta$, as $P_0(\theta,t)=\sum_{l=-\infty}^\infty P_0^{(l)}(t) e^{i l \theta}$, we find on using Eq.~\eqref{eq:FK_diff_ring} that the Fourier coefficients $P_0^{(l)}(t)$ satisfy
\begin{align}
    \frac{d P_0^{(l)}(t)}{ d t} = -D l^2 P_0^{(l)}(t),
\end{align}
which, with the initial condition $P_0(\theta,t=0)=\delta(\theta-\theta_0)$, implying $P_0^{(l)}(0)=(1/(2\pi)) e^{-i l\theta_0}$, has the solution 
\begin{align}
    P_0(\theta,t)=\frac{1}{2 \pi}\sum_{l=-\infty}^\infty  e^{i l(\theta-\theta_0)-l^2 D t}.\label{eq:free_prop_ring}
\end{align}
The above equation also gives the conditional probability density $P_0(\theta,t|\theta_0,0)$ for the particle to be at location $\theta$ at time $t$, given that it was at location $\theta_0$ at time $t=0$. The above equation implies the following stationary-state density, attained in the limit $t \to \infty$:
\begin{align}
    P_{0,\mathrm{ss}}(\theta)=\frac{1}{2\pi}.
    \label{eq:stationary-state-single-particle}
\end{align}
In the stationary state, the particle is equally likely to be found anywhere on the circle. The stationary state~\eqref{eq:stationary-state-single-particle} exemplifies an equilibrium stationary state.

A modified version of dynamics~\eqref{eq:bare-dynamics} corresponds to having an angular frequency term on the right-hand side (RHS):
\begin{align}
    \frac{d\theta}{dt}=\omega+\sqrt{2D}~\eta(t),
    \label{eq:bare-dynamics-1}
\end{align}
which however can be easily reduced to the dynamics~\eqref{eq:bare-dynamics} through the transformation $\theta \to \theta - \omega t$, the latter being tantamount to viewing the dynamics in a co-rotating frame moving with constant angular frequency $\omega$ with respect to an inertial frame. In such a co-rotating frame, the probability density of the particle will be stationary and will be given by Eq.~\eqref{eq:stationary-state-single-particle}. In the inertial frame, the density will satisfy
\begin{align}
    P_\mathrm{inertial}(\theta,t)=P_0(\theta - \omega t,t),
\end{align}
with $P_0(\theta,t)$ given by Eq.~\eqref{eq:free_prop_ring}; the above represents a time-periodic state. We thus conclude that no stationary state is ever reached in the inertial frame.

\subsection{Dynamics in presence of resetting}
\label{sec:single-particle-reset}

We now consider the case of  resetting dynamics of the system~\eqref{eq:bare-dynamics}, which involves the following update rules of the angle $\theta(t)$ in a small time interval $[t,t+dt]$:
\begin{align}
&\theta(t+dt)\nonumber \\
&= 
\begin{cases} 
    \theta_\mathrm{r}=0& \text{with probability } \lambda ~ dt, \\
    \theta(t) + \sqrt{2D ~ dt} ~ \eta(t) & \text{with probability } (1 - \lambda ~ dt),
\end{cases}
\label{eq:global_reset-protocol_diffusive_ring}
\end{align}
where $\lambda > 0$ is the resetting rate, and $\theta_\mathrm{r}$ is a given location, the so-called reset location, on the circle. Thus, the resetting dynamics involves the particle in a small time $[t,t+dt]$ either resetting to location $\theta_\mathrm{r}$ with probability $\lambda dt$ or updating its position on the circle according to the bare dynamics~\eqref{eq:bare-dynamics} with the complementary probability $1-\lambda dt$. The time interval $\tau$ between two successive resets is then a random variable distributed according to an exponential, $p(\tau)=\lambda e^{-\lambda \tau}$. As $\lambda \to 0$, one recovers the bare dynamics~\eqref{eq:bare-dynamics}.

It is straightforward to write down the time evolution equation for the probability density $P_\mathrm{r}(\theta,t)$ in presence of resetting, which reads as
\begin{align}
    &\frac{\partial P_\mathrm{r}(\theta,t)} {\partial t}=D \frac{\partial^2 P_\mathrm{r}(\theta,t)} {\partial \theta^2} -\lambda P_\mathrm{r}(\theta,t)+ \lambda\delta(\theta)\label{master_eq_resetting_ring},
\end{align}
with the initial condition $P_\mathrm{r}(\theta,0)=\delta(\theta-\theta_0)$.
Along with the diffusive spread represented by the first term on the RHS of Eq.~\eqref{master_eq_resetting_ring}, the second term implies loss of probability at any angle $\theta$ due to resetting to angle $\theta_\mathrm{r}$ with rate $\lambda$. The third term implies the increase in probability at $\theta_\mathrm{r}$ due to resetting from all other $\theta$'s with rate $\lambda$. Equivalent to Eq.~\eqref{master_eq_resetting_ring}, one has the following renewal equation for the conditional probability density $P_\mathrm{r}(\theta,t|\theta_0,0)$ in terms of the conditional probability density $P_0(\theta,t|\theta_0,0)$ in absence of resetting given by Eq.~\eqref{eq:free_prop_ring}~\cite{evans2020stochastic}:
\begin{align}
    P_\mathrm{r}(\theta,t|\theta_0,0)&= e^{-\lambda t}  P_0(\theta,t|\theta_0,0)\nonumber \\&+
\lambda \int_0^t d\tau~e^{-\lambda \tau}~P_0(\theta,t|\theta_\mathrm{r},t-\tau)
\label{eq:reset_renewal_ring_1}.
\end{align}
Here, the first term on the RHS represents particular dynamical trajectories connecting initial location $\theta_0$ to the location of interest $\theta$ at time $t$, which did not undergo a single reset during time $t$. On the other hand, the second term represents trajectories that underwent last resetting at time instant $t-\tau$ and reset-free evolution from times $t-\tau$ to $t$, with the variable $\tau$ taking values over the range $[0,t]$.
Using Eq.~\eqref{eq:free_prop_ring}, we get
\begin{align}
P_\mathrm{r}(\theta,t|\theta_0,0)&=\frac{1}{2\pi}\sum_{l=-\infty}^\infty  e^{i l(\theta-\theta_0)}e^{-(\lambda+D l^2) t}\nonumber \\
&+
\frac{\lambda}{2 \pi} \sum_{l=-\infty}^\infty e^{i l\theta}\left( \frac{1-e^{-(\lambda+ D l^2 )t} }{\lambda+D l^2}\right)\label{eq:reset_renewal_ring_2},
\end{align}
which using $\int_{-\pi}^{\pi} d\theta~e^{i (l-m) \theta}= 2\pi \delta_{l,m}$ is easily checked to be normalized. 
\begin{figure}[htbp!]
    \centering
\includegraphics[width=0.7\linewidth,height=0.6\linewidth]{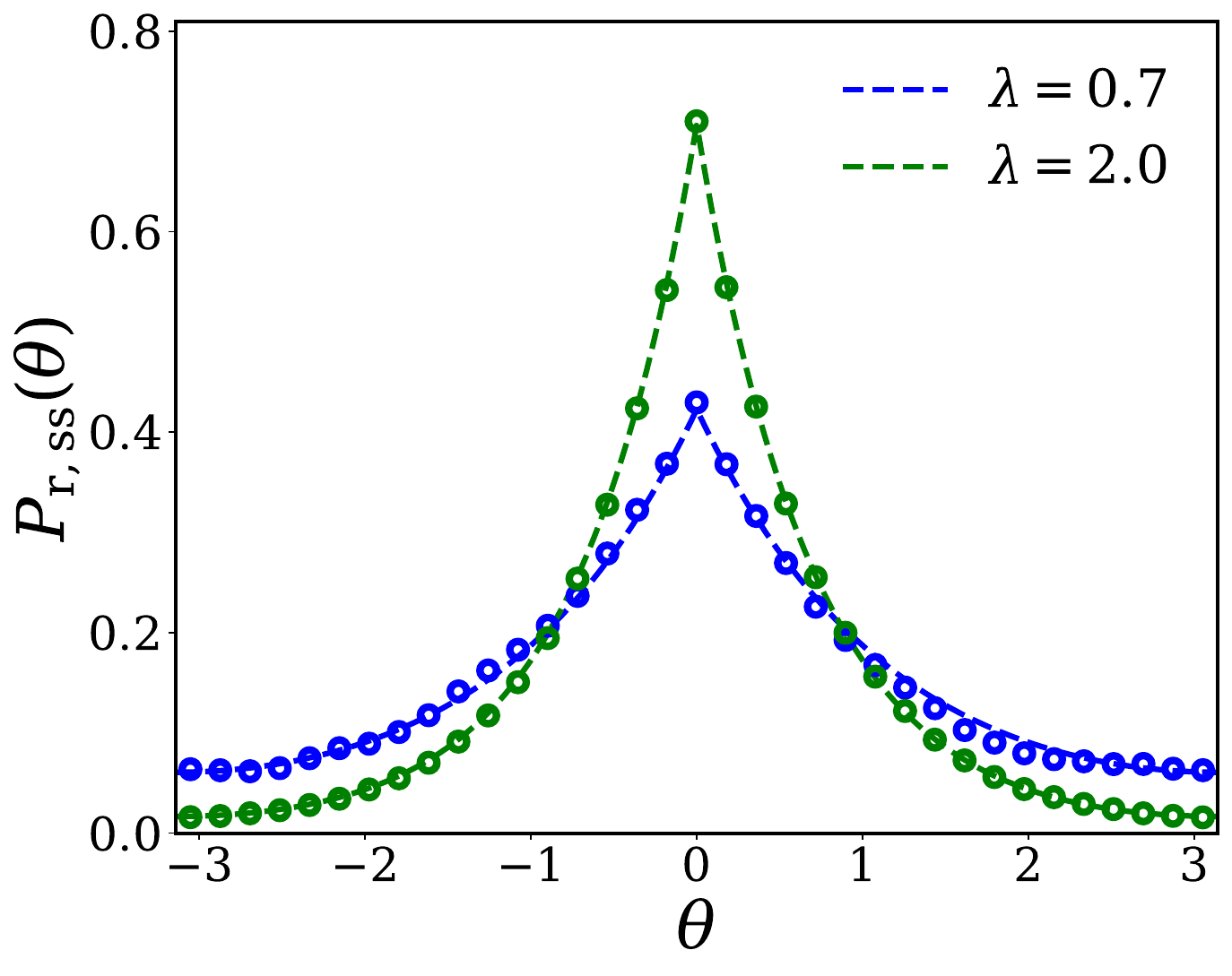}
    \caption{For the problem of a single diffusing particle on a circle subject to reset to a fixed location $\theta_\mathrm{r}=0$, Sec.~\ref{sec:single-particle-reset}, the figure shows match between analytical (Eq.~\eqref{reset_renewal_ring-ss}, dashed lines) and simulation (symbols) results for the stationary-state density, for two resetting rates $\lambda=0.7$ and $\lambda=2.0$.}
    \label{fig:result_Pss-single-particle}
\end{figure}
As $t \to \infty$, Eq.~\eqref{eq:reset_renewal_ring_2} yields the stationary-state density as
\begin{align}
    P_\mathrm{r,ss}(\theta)= \frac{\lambda}{2 \pi} \sum_{l=-\infty}^\infty  \frac{e^{i l\theta}}{\lambda+D l^2}\label{reset_renewal_ring-ss}.
\end{align}

The stationary state in presence of resetting has the particle most likely to be found at the reset location $\theta_\mathrm{r}=0$; this is quite unlike the case in the absence of resetting when the particle is equally likely to be found anywhere on the circle, see Eq.~\eqref{eq:stationary-state-single-particle}. As is well known, a stationary state obtained in presence of resetting manifestly breaks detailed balance and is an NESS~\cite{evans2020stochastic} and hence, the density~\eqref{reset_renewal_ring-ss} represents an NESS. 
In Fig.~\ref{fig:result_Pss-single-particle}, we provide an exact match between analytical and simulation results for the stationary-state density. All numerical simulation results reported in this work corresponding to deterministic dynamics are obtained by employing a standard fourth-order Runge-Kutta integrator with integration time step $=0.01$. The corresponding stochastic dynamics is integrated by using a combination of fourth-order Runge-Kutta and Euler integration schemes with the same choice of the integration step.

\section{Resetting in interacting limit-cycle oscillator systems}
\label{sec:many-particles}

\subsection{Dynamics in absence of resetting}
\label{sec:NK-no-reset}

We now move on to consider a model of interacting limit-cycle oscillator systems. In order to make connection with the preceding section, we first motivate the model from the point of view of interacting particles on a circle. To this end, we consider a collection of $N$ point particles diffusing on a unit circle in the presence of a mean-field inter-particle interaction potential. In addition, every particle has the possibility of an independent motion dictated by its intrinsic angular frequency. Specifically, denoting the angular position of the $i$-th particle, $i=1,2,\ldots, N$, on the circle by $\theta_i \in [-\pi,\pi]$,  its equation of motion is given by
\begin{align}
    \frac{d\theta_i}{dt}=\omega_i+\frac{K}{N}\sum_{j=1}^N \sin(\theta_j-\theta_i)+\sqrt{2D}~\eta_i(t),
    \label{eq:noisy-kuramoto-eom}
\end{align}
where the first term on the RHS denotes the intrinsic angular frequency associated with the $i$-th particle, while the second term is the force (more precisely, the torque) derived from the mean-field potential $(K/N)\sum_{i,j=1}^N [1-\cos(\theta_i-\theta_j)]$, with $K$ being the coupling constant. The Gaussian, white noise $\eta_i(t)$ satisfies
\begin{align}
    \langle \eta_i(t)\rangle=0~\forall~i,~\langle \eta_i(t)\eta_j(t')\rangle=\delta_{ij}\delta(t-t')~\forall~i,j.
    \label{eq:noise-Kuramoto}
\end{align}
Particles with the same value of intrinsic angular frequency $\omega$ in their dynamics are indistinguishable.

\begin{figure}[htbp!]
    \centering
\includegraphics[width=1\linewidth]{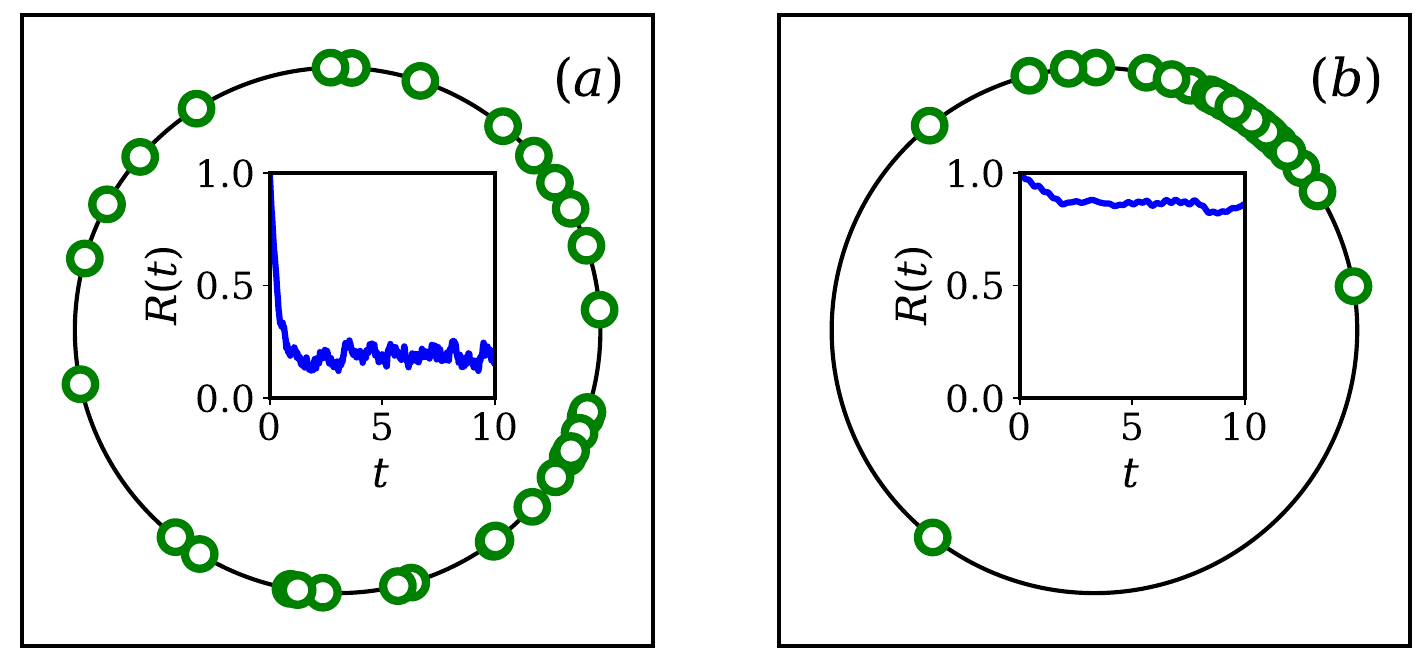}
    \caption{Schematic representation of the quantity $R$, see Eq.~\eqref{eq:r-defn}, for (a) a declustered ($R \approx 0$) and (b) a clustered ($R \approx 1$) state. Each inset shows how declustering/clustering develops in time starting from a fully clustered state with $R=1$ and evolving according to the bare Kuramoto dynamics, Eq.~\eqref{Eq:kuramoto-eom}, with the natural frequencies of the oscillators chosen from the Lorentzian distribution~\eqref{eq:Lorentzian}, and with $\sigma=5.0$ (panel (a)) and $\sigma=0.3$ (panel (b)).}
    \label{fig:R-schematic}
\end{figure}
The mean-field potential promotes clustering of particles on the circle. Clustering at any time $t$ can be quantified in terms of the order parameter~\cite{kuramoto1984chemical}
\begin{align}
    R(t)e^{i\psi(t)}\equiv\frac{1}{N}\sum_{j=1}^N e^{i\theta_j(t)}.
    \label{eq:r-defn}
\end{align}
Here, the vector $\textbf{R}(t)=R(t)e^{i\psi(t)}$, $0 \le R(t) \le 1;~\psi(t) \in [-\pi,\pi]$, measures clustering in terms of its length $R(t)$, while $\psi(t)$ denotes the inclination of the vector with respect to an arbitrary zero-$\theta$ axis; $R=0$ (respectively, $R=1$) denotes a completely declustered (respectively, a completely clustered) state.

Different from the particle-based interpretation of the dynamics~\eqref{eq:noisy-kuramoto-eom}, it may also be taken to describe the time evolution of $N$ coupled limit-cycle oscillators, each characterized by its phase $\theta_i \in [-\pi,\pi]$ and by its intrinsic angular frequency or natural frequency $\omega_i$, which are coupled all-to-all via an interaction that favors synchronization, namely, a state in which the phases evolve in time while maintaining time-independent phase differences. In this interpretation involving limit-cycle oscillators, the quantities $R$ and $\psi$ characterize respectively the amount
of synchronization and the average phase; $0<R<1$ characterizes a synchronized state, while $R=0$ characterizes an incoherent state. The dynamics~\eqref{eq:noisy-kuramoto-eom} is then said to describe the so-called noisy Kuramoto model, while setting $D$ to zero, one obtains the celebrated Kuramoto model discussed in the introduction. The BMF model corresponds to the case $\omega_i=0~\forall~i$. In the following, we will use the particle and the oscillator interpretation interchangeably. Particles with the same value of intrinsic angular frequency being indistinguishable translate to oscillators with the same natural frequency being indistinguishable.

The set of natural frequencies $\{\omega_i;-\infty<\omega_i<\infty\}_{1\le i \le N}$ denote $N$ quenched-disordered random variables sampled independently from a common distribution $g(\omega)$ with mean $\omega_0$ and finite width $\sigma>0$. The latter may be taken to be the standard deviation associated with $g(\omega)$ when it is finite and is otherwise taken to be the full-width-at-half-maximum of $g(\omega)$. The Gaussian, white noise $\eta_i(t)$ models annealed, i.e., time-dependent disorder in the dynamics. 

As is usual in studies of the Kuramoto model, we will take $g(\omega)$ to be unimodal, i.e., one which is symmetric about the mean $\omega_0$, and which decreases monotonically and continuously to zero with increasing $|\omega-\omega_0|$ (in one particular case of study, we will take $g(\omega)$ to be a uniform distribution, which is symmetric about the mean $\omega_0$, but which does not decay with increasing $|\omega-\omega_0|$.). One may get rid of the effect of the mean $\omega_0$ from the dynamics~\eqref{eq:noisy-kuramoto-eom} by viewing it in a frame co-rotating with constant angular frequency $\omega_0$ with respect to an inertial frame. Furthermore, the effect of $\sigma$ in the dynamics may be made explicit by scaling the frequency variables as $\omega_i \to \omega_i/\sigma$, so that the dynamics reads as 
\begin{align}
    \frac{d\theta_i}{dt}=\sigma\omega_i+\frac{K}{N}\sum_{j=1}^N \sin(\theta_j-\theta_i)+\sqrt{2D}~\eta_i(t),
    \label{eq:Noisy-Kuramoto-eom-1}
\end{align}
where the $\omega_i$'s are now quenched-disordered random variables distributed according to a distribution $g(\omega)$ that has mean zero and width equal to unity; in particular, $g(\omega)=g(-\omega)$, that is, any frequency value $\omega$ and the corresponding negative value $-\omega$ have the same probability of occurrence.

The dynamics dictated by Eq.~\eqref{eq:Noisy-Kuramoto-eom-1} has associated with it three dynamical parameters $\sigma,K, D$, one of which may be gotten rid of from the dynamics through a suitable scaling of variables. For example, let us note the dimension of the following variables: $[\sigma]=1/T$, $[K]=1/T$, $[D]=1/T$, where $T$ denotes time dimension. This suggests the following scaling of the variables: $\sigma \to \sigma/K$, $D \to D/K$ and $t \to Kt$. The dynamics~\eqref{eq:Noisy-Kuramoto-eom-1} in terms of scaled variables writes as 
\begin{align}
    \frac{d\theta_i}{dt}=\sigma\omega_i+\frac{1}{N}\sum_{j=1}^N \sin(\theta_j-\theta_i)+\sqrt{2D}~\eta_i(t),
    \label{eq:NK-eom-1}
\end{align}
with the noise $\eta_i(t)$ satisfying the same properties as in Eq.~\eqref{eq:noise-Kuramoto}.  In the above equation, all quantities are dimensionless.

The dynamics~\eqref{eq:NK-eom-1} can be characterized in terms of the single-particle probability density $P(\theta,\omega,t)$, which has the interpretation that $P(\theta,\omega,t)d\theta$ gives out of all oscillators with frequency $\omega$ the fraction that has the phase between $\theta$ and $\theta+d\theta$ at time $t$. This density is normalized as $\int_{-\pi}^{\pi}d\theta~P(\theta,\omega,t)=1~\forall~\omega,t$, and is $2\pi$-periodic in $\theta$. Its time evolution may be obtained by considering its change in a small time $[t,t+dt]$:
\begin{widetext}
\begin{align}
    &P(\theta,\omega,t+dt)=\langle P(\theta-dt~(\sigma \omega+\int d\theta'd\omega'~g(\omega')P(\theta',\omega'|\theta,\omega,t)\sin(\theta'-\theta))-\sqrt{2D~dt}~\eta(t),\omega,t)\rangle_{\eta(t)},
\label{eq:NK_probability_dynamics}
\end{align}
\end{widetext}
where $P(\theta',\omega'|\theta,\omega,t)$ is the conditional probability density. We invoke a mean-field approximation, implying the following
factorization property of the joint probability density:
\begin{align}
P_2((\theta,\omega),(\theta',\omega'),t)=P(\theta,\omega,t)P(\theta',\omega',t), 
\label{eq:P2}
\end{align} 
implying 
\begin{align}
P(\theta',\omega'|\theta,\omega,t)=P(\theta',\omega',t).
\label{eq:factorization}
\end{align}
The mean-field approximation encodes the fact that particles in different oscillator frequency groups are uncorrelated in the sense that the event that one finds an oscillator with phase $\theta$ in the group of oscillators with frequency $\omega$ is independent of the event of finding an oscillator with phase $\theta'$ in the group of oscillators with a different frequency $\omega'$. 
We have the normalization $\int d\theta d\theta'~P_2((\theta,\omega),(\theta',\omega'),t)=1~\forall~\omega,\omega',t$ and $\int d\theta'~P(\theta',\omega'|\theta,\omega,t)=1~\forall~\theta,\omega,t$. The factorization in Eq.~\eqref{eq:P2} on using in Eq.~\eqref{eq:NK_probability_dynamics} leads to the equation
\begin{align}
    &P(\theta,\omega,t+dt)=\nonumber \\
    &\langle P(\theta-dt~\left(\sigma \omega+R(t)\sin\left(\psi(t)-\theta\right)\right)-\sqrt{2D~dt}~\eta(t),t)\rangle_{\eta(t)},
\label{eq2:NK_probability_dynamics}
\end{align}
where we have
\begin{align}
    R(t)e^{i\psi(t)}&=\int d\theta d\omega~e^{i\theta}P(\theta,\omega,t)g(\omega)\nonumber \\
    &\equiv \int d\omega~g(\omega)\langle e^{i\theta}\rangle(\omega,t),
    \label{eq:r-eqn-continuum-NK}
\end{align}
which follows from Eq.~\eqref{eq:r-defn} in the limit $N \to \infty$. Note that the function $P_2((\theta,\omega),(\theta',\omega'),t)$ is symmetric with respect to permutations between groups of oscillators with the same frequency,
\begin{align}
P_2((\theta,\omega),(\theta',\omega'),t)=P_2((\theta',\omega'),(\theta,\omega),t).
\label{eq:P2-symmetry}
\end{align}
This is because oscillators with the same frequency are indistinguishable.
In the limit $dt \to 0$, Eq.~\eqref{eq2:NK_probability_dynamics} yields the Fokker-Planck equation
\begin{align}
    \frac{\partial P(\theta,\omega,t)}{\partial t}&=-\frac{\partial[(\sigma \omega+R(t)\sin(\psi(t)-\theta))P(\theta,\omega,t)]}{\partial \theta}\nonumber \\
    &+D\frac{\partial^2 P(\theta,\omega,t)}{\partial \theta^2}.
    \label{eq:NK_FP}
\end{align}

In the stationary state, attained as $t\to \infty$, both $R$ and $\psi$ become time-independent. The stationary value of $\psi$ can be set to zero by redefining the zero-$\theta$ axis. Consequently, the stationary-state density $P_\mathrm{ss}(\theta,\omega)$, which characterizes an NESS~\cite{gupta2018statistical}, satisfies 
\begin{align}
    &-\frac{\partial[(\sigma \omega-R_\mathrm{ss}\sin \theta)P_\mathrm{ss}(\theta,\omega)]}{\partial \theta}+D\frac{\partial^2 P_\mathrm{ss}(\theta,\omega)}{\partial \theta^2}=0,
    \label{eq:NK-FP-ss}
\end{align}
where the equation satisfied by the stationary value of $R$, denoted by $R_{\mathrm{ss}}$ may be found as follows: We start with the stationary-state form of Eq.~\eqref{eq:r-eqn-continuum-NK}, which with stationary $\psi=0$ reads as
\begin{align}
 R_\mathrm{ss}&=\int d\theta d\omega~e^{i\theta}P_\mathrm{ss}(\theta,\omega)g(\omega).   
\end{align}
Equating the real and the imaginary part on both sides of the above equation, we get
\begin{align}
    &R_\mathrm{ss}=\int d\theta d\omega~\cos \theta~P_\mathrm{ss}(\theta,\omega)g(\omega),\nonumber \\
    \label{eq:r-eqn-continuum-NK-ss} \\
    &0=\int d\theta d\omega~\sin \theta~P_\mathrm{ss}(\theta,\omega)g(\omega). \nonumber
    \end{align}
The last equation is automatically satisfied because $P_\mathrm{ss}(\theta,\omega)=P_\mathrm{ss}(-\theta,-\omega)~\forall~\omega$, as follows from Eq.~\eqref{eq:NK-FP-ss} on using the fact that $g(\omega)=g(-\omega)$.

\subsection{Dynamics in presence of resetting}
\label{sec:NK-reset}

In analogy with the reset dynamics of the single particle on a circle, Eq.~\eqref{eq:global_reset-protocol_diffusive_ring}, we now study the resetting dynamics of the system~\eqref{eq:NK-eom-1}, which involves the following update rules of the angles $\theta_i(t)~\forall~i$ in a small time interval $[t,t+dt]$:
\begin{align}
\theta_i(t+dt)=
\begin{cases}
               &\theta_\mathrm{r}=0 \mbox{~with~probability~}\lambda~dt, \\ \\
               &\theta_i(t)\nonumber \\
               &+[\sigma \omega_i+\frac{1}{N}\sum_{j=1}^N\sin(\theta_j(t)-\theta_i(t))]dt\nonumber \\
               &+\sqrt{2D ~dt}~\eta_i(t) \\
               & \mbox{with~probability~}(1-\lambda ~dt). 
         \end{cases} \\
\label{eq:global_reset-protocol_noisyKuramoto}
\end{align}

Thus, in the resetting protocol, the particles all together undergo resetting to the common location $\theta_\mathrm{r}$. This corresponds to the order parameter $R$ resetting to the value of unity. In Appendix~\ref{sec2:app1}, we discuss the case when the order parameter is reset to any value smaller than unity. 

The equivalent of Eq.~\eqref{eq:NK_probability_dynamics} for the case at hand reads as
\begin{widetext}
\begin{align}
    P_\mathrm{r}(\theta,\omega,t+dt)&=(1-\lambda dt)\langle P_\mathrm{r}(\theta-dt~(\sigma \omega+\int d\theta'd\omega'g(\omega')~P_\mathrm{r}(\theta',\omega'|\theta,\omega,t)\sin(\theta'-\theta))-\sqrt{2D~dt}~\eta(t),\omega,t)\rangle_{\eta(t)}\nonumber \\
    &+\lambda dt~\delta(\theta).
\label{eq:NK-reset}
\end{align}
\end{widetext}
One may understand the above equation as follows. To contribute to $P_\mathrm{r}(\theta,\omega,t+dt)$, one would have to consider the fraction of the particles with frequency $\omega$ that were at location $\theta-dt~(\sigma \omega+\int d\theta'd\omega'~P_\mathrm{r}(\theta',\omega'|\theta,\omega,t)\sin(\theta'-\theta))-\sqrt{2D~dt}~\eta(t)$ at time $t$ and which underwent no reset in time $[t,t+dt]$, or, in case we are considering $\theta=0$, one would have to consider a reset to $\theta_\mathrm{r}=0$. The first contribution involves the stochastic noise $\eta(t)$, and thus one has to average over all possible values of $\eta(t)$, represented by the angular brackets in the first term on the RHS of Eq.~\eqref{eq:NK-reset}. 

To proceed, let us implement factorization of the sort in Eq.~\eqref{eq:P2} leading to a result of the form~\eqref{eq:factorization}:
\begin{align}
P_\mathrm{r}(\theta',\omega'|\theta,\omega,t)=P_\mathrm{r}(\theta',\omega',t).
\label{eq:factorization-1}
\end{align}
Note that Eq.~\eqref{eq:P2-symmetry} will also hold in presence of resetting, since under resetting, all the oscillator phases are reset to the common value $\theta_\mathrm{r}=0$:
\begin{align}
P_{2,\mathrm{r}}((\theta,\omega),(\theta',\omega'),t)=P_{2,\mathrm{r}}((\theta',\omega'),(\theta,\omega),t).
\label{eq:P2-symmetry-1}
\end{align}
Using Eq.~\eqref{eq:factorization-1}, in the limit $dt \to 0$, we get from Eq.~\eqref{eq:NK-reset} the equation 
\begin{align}
    \frac{\partial P_\mathrm{r}(\theta,\omega,t)}{\partial t}&=-\frac{\partial [(\sigma \omega+R(t)\sin(\psi(t)-\theta)P_\mathrm{r}(\theta,\omega,t)]}{\partial \theta}\nonumber \\
    &+D\frac{\partial^2 P_\mathrm{r}(\theta,\omega,t)}{\partial \theta^2}-\lambda P_\mathrm{r}(\theta,\omega,t)+\lambda\delta(\theta),\label{eq:NK-reset-FP_dyn}
\end{align}
with $R(t)$ and $\psi(t)$ given by Eq.~\eqref{eq:r-eqn-continuum-NK} with $P(\theta,\omega,t)$ replaced by $P_\mathrm{r}(\theta,\omega,t)$.
Then, the stationary-state density satisfies the equation
\begin{align}
    &-\frac{\partial[(\sigma \omega-R_\mathrm{ss}\sin \theta)P_\mathrm{r,ss}(\theta,\omega)]}{\partial \theta}+D\frac{\partial^2 P_\mathrm{r,ss}(\theta,\omega)}{\partial \theta^2}\nonumber \\
    &-\lambda P_\mathrm{r,ss}(\theta,\omega)+\lambda \delta(\theta)=0,
    \label{eq:NK-reset-FP-ss}
\end{align}
with $R_\mathrm{ss}$ satisfying Eq.~\eqref{eq:r-eqn-continuum-NK-ss} with $P_\mathrm{ss}(\theta,\omega)$ replaced by $P_\mathrm{r,ss}(\theta,\omega)$. Setting $\lambda$ to zero (the no-resetting case) in Eq.~\eqref{eq:NK-reset-FP-ss} correctly recovers the reset-free stationary-state equation~\eqref{eq:NK-FP-ss}.

In the following sections, we will discuss the behavior of the stationary state across several scenarios by considering Eq.~\eqref{eq:NK-reset-FP-ss}, both in the absence and presence of resetting. First, we will consider the case in which the particles have no intrinsic angular frequency in their dynamics (the case $\sigma=0$, defining the BMF model). Next, we will analyze the case with frequencies ($\sigma \neq 0)$, considering in particular Lorentzian and uniform distributions for the natural frequencies; here, we will take up both the noiseless case, defining the bare Kuramoto model, and the noisy case corresponding to the noisy Kuramoto model. To proceed, we discuss in the next subsection a general method to obtain the stationary state from Eq.~\eqref{eq:NK-reset-FP-ss}.

\subsection{General approach for obtaining the stationary state}
\label{sec:general-method}

In the following, we develop an analytical approach to obtain the stationary-state density $P_\mathrm{r,ss}(\theta,\omega)$ by solving Eq.~\eqref{eq:NK-reset-FP-ss}. This method addresses collectively all the dynamical scenarios mentioned in the concluding part of the previous subsection. To this end, we express $P_\mathrm{r,ss}(\theta,\omega)$ as a Fourier series:
 \begin{align}
    P_\mathrm{r,ss}(\theta,\omega)=\sum_{l=-\infty}^{\infty} P_\mathrm{r,ss}^{(l)}(\omega) e^{i l \theta},
    \label{eq:NK-Fourier}
 \end{align}
  with $P_\mathrm{r,ss}(\theta,\omega)$ being real implying that the Fourier coefficients satisfy $P_\mathrm{r,ss}^{(-l)}(\omega)=[P_\mathrm{r,ss}^{(l)}(\omega)]^\star$ for all $\omega$ and with the star denoting complex conjugation. Additionally, the Dirac-delta function  
  can be expanded as $\delta(\theta)=(1/(2\pi)) \sum_{l=-\infty}^{\infty} e^{i l \theta}$. Substituting these expansions in Eq.~\eqref{eq:NK-reset-FP-ss}, we obtain the following recurrence relation for $P_\mathrm{r,ss}^{(l)}(\omega)$:
\begin{align}
&b~l P_\mathrm{r,ss}^{(l+1)}+(D l^2+i \sigma \omega l+\lambda) P_\mathrm{r,ss}^{(l)} - b~l P_{\mathrm{r,ss}}^{(l-1)}-c=0,
\label{eq:recurrence_global_reset_general}
\end{align}
where we have defined 
\begin{align}
b\equiv \frac{R_\mathrm{ss}}{2},~c\equiv\frac{\lambda}{2 \pi}.
\label{eq:bc-defn}
\end{align}
Equation~\eqref{eq:recurrence_global_reset_general} holds for all $l$ and implies for $\sigma=0$ that $P_\mathrm{r,ss}^{(-l)}=[P_\mathrm{r,ss}^{(l)}]^\star=P_\mathrm{r,ss}^{(l)}$.
Using the normalization, $\int_{-\pi}^\pi d\theta~P_\mathrm{r,ss}(\theta,\omega)=1~\forall~\omega$, it follows that
\begin{align}
P_\mathrm{r,ss}^{(0)}(\omega)=\frac{1}{2 \pi}~\forall~\omega. 
\label{eq:zero-Fourier-coefficient}
\end{align} 
On the other hand, Eq.~\eqref{eq:r-eqn-continuum-NK-ss}, with $P_\mathrm{r,ss}$ replacing $P_\mathrm{ss}$ and use of Eq.~\eqref{eq:NK-Fourier}, yields
\begin{align}
    R_\mathrm{ss}&=2 \pi\int  d\omega~g(\omega)~\mathrm{Re}[P_\mathrm{r,ss}^{(-1)}(\omega)]\nonumber\\
    &=2 \pi\int  d\omega~g(\omega)~\mathrm{Re}[P_\mathrm{r,ss}^{(1)}(\omega)],\label{eq:rss_in_terms_of_P_{r,ss}^1}
\end{align}
where in the last step, we have used $P_\mathrm{r,ss}^{(-1)}=[P_\mathrm{r,ss}^{(1)}]{}^*$ and $\mathrm{Re}[P_\mathrm{r,ss}^{(-1)}]=\mathrm{Re}[[P_\mathrm{r,ss}^{(1)}]^*]=\mathrm{Re}[P_\mathrm{r,ss}^{(1)}]$. Here, Re stands for ``real part of." 

Let us first consider the case $b=0$, i.e., $R_\mathrm{ss}=0$. In this case, Eq.~\eqref{eq:recurrence_global_reset_general} gives 
\begin{align}
&P_\mathrm{r,ss}^{(l)}(\omega)=\frac{\lambda}{(D l^2+i \sigma \omega l+\lambda)2\pi}, 
\label{eq:recurrence_global_reset_general-b0}\end{align}
implying 
\begin{align}
\mathrm{Re}[P_\mathrm{r,ss}^{(1)}(\omega)]=\frac{\lambda(D+\lambda)}{((D+\lambda)^2+\sigma^2\omega^2)2\pi}.
 \end{align}
For $\lambda=0$, regardless of the value of $D$ and $\sigma$, the above expression when used in Eq.~\eqref{eq:rss_in_terms_of_P_{r,ss}^1} consistently gives $R_\mathrm{ss}=0$. On the other hand, following the same step for $\lambda\ne0$ does not yield consistently that $R_\mathrm{ss}=0$. We may therefore conclude that the dynamics~\eqref{eq:NK-eom-1} admits a completely declustered/incoherent stationary state, while adding resetting to the dynamics, as in Eq.~\eqref{eq:global_reset-protocol_noisyKuramoto}, never allows such a state. 

We now move on to consider $b\ne0$, i.e., the case with a synchronized/clustered stationary state. With the aim to obtain the value of the corresponding $R_\mathrm{ss}$, our main goal now is to evaluate $P_{\mathrm{r,ss}}^{(1)}(\omega)$ from Eq.~\eqref{eq:recurrence_global_reset_general}. From this equation, it can be checked that $P_{\mathrm{r,ss}}^{(2)}$ depends on $P_\mathrm{r,ss}^{(1)}$ and $P_\mathrm{r,ss}^{(0)}$, while $P_{\mathrm{r,ss}}^{(3)}$ depends upon $P_\mathrm{r,ss}^{(2)}$ and $P_\mathrm{r,ss}^{(1)}$ and therefore on $P_\mathrm{r,ss}^{(2)}$ and $P_\mathrm{r,ss}^{(0)}$; note that $P_\mathrm{r,ss}^{(0)}$ is already known and is given by Eq.~\eqref{eq:zero-Fourier-coefficient}. This suggests to make the ansatz~\cite{risken1980solutions}
\begin{align}
P_{\mathrm{r,ss}}^{(l)}(\omega)= S^{(l)} P_{\mathrm{r,ss}}^{(l-1)}(\omega)+ \beta^{(l)};~l \ne 0, 
\label{eq:the_ansatz_general}
\end{align}
with the quantities $S^{(l)}$ and $\beta^{(l)}$ to be determined.
Using the ansatz~\eqref{eq:the_ansatz_general} in Eq.~\eqref{eq:recurrence_global_reset_general}, we get
\begin{align}
    P_{\mathrm{r,ss}}^{(l)}(\omega)= \frac{b~l}{ G^{(l)}} P_{\mathrm{r,ss}}^{(l-1)}(\omega)+\frac{c- b~l \beta^{(l+1)}}{G^{(l)}};~l \ne 0,
    \label{eq:Prss-Glbetal}
\end{align}
with $G^{(l)} \equiv b~l S^{(l+1)} +(D l^2+i \sigma \omega l+\lambda)$. A comparison with Eq.~\eqref{eq:the_ansatz_general} then yields the following recurrence relation for the quantities $S^{(l)}$ and $\beta^{(l)}$ for $l \ne 0$:
\begin{align}
    &S^{(l)}=\frac{b~l}{b~l S^{(l+1)} +(D l^2+i \sigma\omega l+\lambda)},\nonumber \\
    \label{eq:continued_fraction_general}\\
    &\beta^{(l)}=\frac{c- b~l \beta^{(l+1)}}{b~l S^{(l+1)} +(D l^2+i \sigma\omega l
    +\lambda)}. \nonumber
\end{align}

 In Appendix~\ref{subsec:app1}, we study the large-$l$ behavior of $S^{(l)}$ and $\beta^{(l)}$, based on which
 we conclude that as long as $D \ne 0$, the quantities $S^{(l)}$ and $\beta^{(l)}$ for a given $\omega$ decrease with increasing $l$, decaying to zero for large $l$.  Note that the analysis in Appendix~\ref{subsec:app1} does not give the value of $\mathcal{N}$, but only elaborates on the large-$l$ behaviour of $S^{(l)}$. Note further that  for $D=\lambda=0$ (the Kuramoto model in absence of resetting, discussed in Section~\ref{sec:Kuramoto-no-reset}), when one has $S^{(l)}$ independent of $l$ and $\beta^{(l)}=0~\forall~l$, no such truncation of the infinite series in Eq.~\eqref{eq:NK-Fourier} is required and one can in fact sum the infinite series exactly. Another case where no truncation is required is for $\sigma=\lambda=0$ (the BMF model in absence of resetting, discussed in Section~\ref{sec:BMF-no-reset}), where we use directly Eq.~\eqref{eq:recurrence_global_reset_general}.
 
 Putting $l=\mathcal{N}$ in Eq.~\eqref{eq:recurrence_global_reset_general} and considering that $S^{(\mathcal{N}+1)}=0=\beta^{(\mathcal{N}+1)}$ implies that $P_\mathrm{r,ss}^{(\mathcal{N}+1)}=0$, we get
\begin{align}
P_\mathrm{r,ss}^{(\mathcal{N})}&=\frac{b \mathcal{N}}{D \mathcal{N}^2+i \sigma \omega \mathcal{N}+\lambda}P_{\mathrm{r,ss}}^{(\mathcal{N}-1)}\nonumber \\
&+ \frac{c}{D \mathcal{N}^2+i \sigma \omega \mathcal{N}+\lambda},
\end{align}
which gives 
\begin{align}
 &S^{(\mathcal{N})}=\frac{b \mathcal{N}}{D \mathcal{N}^2+i \sigma \omega \mathcal{N}+\lambda}, \nonumber \\ \label{eq:SNbetaN-main} \\
 &\beta^{(\mathcal{N})}=\frac{c}{D \mathcal{N}^2+i \sigma \omega \mathcal{N}+\lambda}. \nonumber   
\end{align}
Knowing $S^{(\mathcal{N})}$ and $\beta^{(\mathcal{N})}$, one may use Eq.~\eqref{eq:continued_fraction_general} to obtain $S^{(l)}$ and $\beta^{(l)}$ for $0<l<\mathcal{N}$, which finally allows to obtain $P_\mathrm{r,ss}^{(1)}$. Note that following this procedure, one obtains $P_\mathrm{r,ss}^{(1)}$ in terms of the quantity $b$, or, in other words, in terms of the quantity $R_\mathrm{ss}$, see Eq.~\eqref{eq:bc-defn}. Such a solution when inserted on the RHS of Eq.~\eqref{eq:rss_in_terms_of_P_{r,ss}^1} yields a self-consistent equation for $R_\mathrm{ss}$, which when solved finally gives the value of $R_\mathrm{ss}$ for given values of $D \ne0$, and for arbitrary values of $\sigma,\lambda$ and any choice of $g(\omega)$.

\section{Application to the BMF model}
\label{sec:BMF}
\subsection{Dynamics in absence of resetting}
\label{sec:BMF-no-reset}
With the general formalism developed in the previous section, we now proceed to apply it to different model systems to study the effects of stochastic resetting. Specifically, we study both the bare dynamics and the dynamics in the presence of resetting. We first consider the BMF model, which, as discussed in Sec.~\ref{sec:NK-reset}, involves a collection of $N$ point particles diffusing on a unit circle in the presence of a mean-field inter-particle interaction potential, but without the intrinsic angular frequency associated with each particle. The corresponding evolution equation is obtained from Eq.~\eqref{eq:Noisy-Kuramoto-eom-1} by putting $\sigma=0$. Proceeding as in Sec.~\ref{sec:NK-no-reset}, we obtain the evolution in terms of dimensionless parameters as  
\begin{align}
    \frac{d\theta_i}{dt}=\frac{1}{N}\sum_{j=1}^N \sin(\theta_j-\theta_i)+\sqrt{2D}~\eta_i(t).
    \label{eq:BMF-eom}
\end{align}
The stationary-state density, now denoted by $P_\mathrm{ss}(\theta)$, satisfies the following equation (see Eqs.~\eqref{eq:NK-FP-ss} and~\eqref{eq:r-eqn-continuum-NK-ss} in which one has to consider $\sigma=0$ and $g(\omega)=\delta(\omega)$, so that one has $P_\mathrm{ss}(\theta)\equiv P_\mathrm{ss}(\theta,\omega=0)$):
\begin{align}
\frac{\partial[R_\mathrm{ss}\sin \theta~P_\mathrm{ss}(\theta)]}{\partial \theta}+D\frac{\partial^2 P_\mathrm{ss}(\theta)}{\partial \theta^2}=0,
\label{eq:FK-BMF-stationary}
\end{align}
with
\begin{align}
    &R_\mathrm{ss}=\int d\theta~\cos \theta~P_\mathrm{ss}(\theta),\nonumber \\
    \label{eq:r-eqn-continuum-BMF-ss} \\
    &0=\int d\theta~\sin \theta~P_\mathrm{ss}(\theta). \nonumber
    \end{align}
The last equation above is satisfied, given that $P_\mathrm{ss}(\theta)=P_\mathrm{ss}(-\theta)$, as follows from Eq.~\eqref{eq:FK-BMF-stationary}.
    
It is easily checked that the equilibrium stationary state
\begin{align}
    P_\mathrm{ss}(\theta)=\frac{e^{(R_\mathrm{ss}/D)\cos \theta}}{\int_{-\pi}^{\pi}d\theta~e^{(R_\mathrm{ss}/D)\cos \theta}}
    \label{eq:BMF-stationary-state}
\end{align}
solves Eq.~\eqref{eq:FK-BMF-stationary}. The self-consistent equation for $R_\mathrm{ss}$, obtained from Eq.~\eqref{eq:r-eqn-continuum-BMF-ss}, reads as~\cite{chavanis2014brownian,gupta2018statistical} 
\begin{align}
    R_\mathrm{ss}=\frac{\int_{-\pi}^{\pi}d\theta~\cos \theta~ e^{(R_\mathrm{ss}/D)\cos \theta}}{\int_{-\pi}^{\pi}d\theta~e^{(R_\mathrm{ss}/D)\cos \theta}}=\frac{I_1(R_\mathrm{ss}/D)}{I_0(R_\mathrm{ss}/D)},
    \label{eq:BMF-rss-consistent}
\end{align}
where $I_n(x)$ is the $n$-th order modified Bessel function of the first kind; specifically, $I_0(x)=\int_{-\pi}^{\pi}d\theta~\exp(x\cos \theta)$ and $I_1(x)=dI_0(x)/dx$. On analyzing Eq.~\eqref{eq:BMF-rss-consistent}, one finds that the system~\eqref{eq:BMF-eom} exhibits a continuous phase transition between a low-$D$ clustered stationary state ($0<R_\mathrm{ss}\le 1$) and a high-$D$ declustered stationary state ($R_\mathrm{ss}=0$) at the critical noise strength $D_c=1/2$~\cite{gupta2018statistical}; see Fig.~\ref{fig:BMF_globalresettting}.

Note that the solution in Eq.~\eqref{eq:BMF-rss-consistent} may also be obtained from the recursion in Eq.~\eqref{eq:recurrence_global_reset_general} by putting $\sigma=\lambda=0$ and making the obvious substitution $P_\mathrm{r,ss}^{(l)} \to P_\mathrm{ss}^{(l)}$, where the $P_\mathrm{ss}^{(l)}$'s are independent of $\omega$, yielding
\begin{align}
P_\mathrm{ss}^{(l+1)}=P_\mathrm{ss}^{(l-1)}-a l P_\mathrm{ss}^{(l)};~~a\equiv \frac{2D}{R_\mathrm{ss}}, \,\, l \neq0. \label{eq:recurrence_bare_BMF}
\end{align}
Equation~\eqref{eq:rss_in_terms_of_P_{r,ss}^1} now yields
\begin{align}
    R_\mathrm{ss} = 2\pi~\mathrm{Re}[P_\mathrm{ss}^{(1)}].\label{eq:rss_BMF_lambda}
\end{align} 
Defining the generating function  
\begin{align}
    \mathbb{P}(z)\equiv \sum_{l=-\infty}^{\infty} P_\mathrm{ss}^{(l)} z^l,
    \label{eq:def_GF}
\end{align}
one obtains from Eq.~\eqref{eq:recurrence_bare_BMF} a first-order linear differential equation for $\mathbb{P} (z)$ given by
\begin{align}
   a z^2 \frac{d \mathbb{P}(z)}{d z}-(z^2-1)\mathbb{P}(z)=0,\label{GF_ODE_bareBMF}
\end{align}
where we have used $P_{\rm ss}^{(-1)}=P_{\rm ss}^{(1)}$. Equation~\eqref{GF_ODE_bareBMF} can be solved to get
\begin{align}
    \mathbb{P}(z)=C~e^{\frac{1}{a}\left(z+\frac{1}{z}\right)}=C\sum_{l=-\infty}^{\infty} I_l\left(\frac{2}{a}\right)z^l,
    \label{eq:GF_ODE_solution_bareBMF}
\end{align}
where $C$ is a constant. In arriving at the last equality above, we have used the fact that the generating function of the modified Bessel function of the first kind is given by~\cite{NIST:DLMF} 
\begin{align}
    e^{\frac{1}{a}(z+\frac{1}{z})}=\sum_{n=-\infty}^{\infty} I_n\left(\frac{2}{a}\right)z^n.
    \label{eq:generating_func}
\end{align}
Then, on using Eqs.~\eqref{eq:def_GF} and~\eqref{eq:GF_ODE_solution_bareBMF}, and the generalization of Eq.~\eqref{eq:zero-Fourier-coefficient} for the case at hand, we obtain
\begin{align}
P_\mathrm{ss}^{(0)}=CI_0(R_\mathrm{ss}/D)=1/(2\pi),~~P_\mathrm{ss}^{(1)}=CI_1(R_\mathrm{ss}/D),
\end{align}
which together with Eq.~\eqref{eq:rss_BMF_lambda} directly yields the result~\eqref{eq:BMF-rss-consistent}. 

\subsection{Dynamics in presence of resetting}
\label{sec:BMF-reset}
The dynamics of the BMF model in presence of resetting follows Eq.~\eqref{eq:global_reset-protocol_noisyKuramoto} with $\sigma=0$. The stationary-state density $P_\mathrm{r,ss}(\theta)$ may be obtained from Eq.~\eqref{eq:NK-reset-FP-ss} as satisfying
\begin{align}
   \frac{\partial [R_\mathrm{ss}\sin \theta~P_\mathrm{r,ss}]}{\partial \theta}+D\frac{\partial^2 P_\mathrm{r,ss}}{\partial \theta^2}-\lambda P_\mathrm{r,ss}+\lambda\delta(\theta)=0
    \label{Eq:stationary_state_BMF-reset},
\end{align}
with $R_\mathrm{ss}$ satisfying Eq.~\eqref{eq:r-eqn-continuum-BMF-ss} with $P_\mathrm{ss}(\theta)$ replaced by $P_\mathrm{r,ss}(\theta)$.
\begin{figure}[htbp!]
    \centering
\includegraphics[width=0.8\linewidth,height=0.6\linewidth]{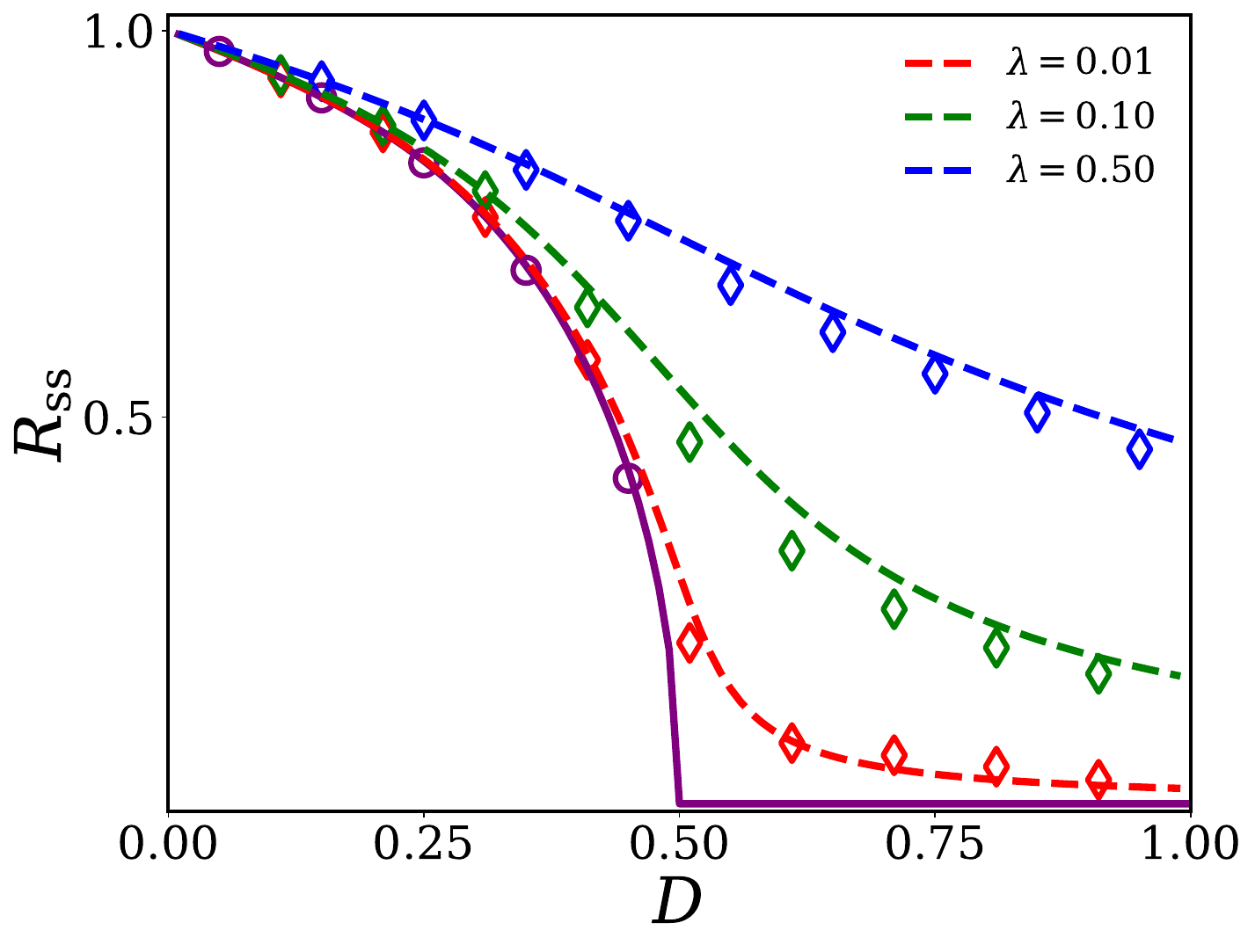}
    \caption{For the BMF model discussed in Sec.~\ref{sec:BMF}, the figure shows the behavior of  $R_\mathrm{ss}$ versus $D$ in the absence (continuous line) and presence (dashed line) of resetting. The former is obtained from analyzing Eq.~\eqref{eq:BMF-rss-consistent}, while the latter is obtained on the basis of analysis in Section~\ref{sec:BMF-reset}. The reset rates are chosen to be $\lambda = 0.01$, $0.10$, and $0.50$.  Simulation results are shown with unfilled diamonds for $\lambda \neq 0$ and unfilled circles for the bare model, both obtained using $N = 10^4$ oscillators; the data correspond to averaging over $50$ dynamical realizations.}
    \label{fig:BMF_globalresettting}
\end{figure}
We now solve Eq.~\eqref{Eq:stationary_state_BMF-reset} by using the method introduced in Sec.~\ref{sec:general-method}, and considering $D \neq 0$, $\lambda \neq 0$, and $\sigma = 0$. Truncation at $l=\mathcal{N}$ of the infinite series in Eq.~\eqref{eq:NK-Fourier}, now expressed for $P_\mathrm{r,ss}(\theta)$, along with Eq.~\eqref{eq:SNbetaN-main} for $\sigma=0$, allows us to obtain $P_\mathrm{r,ss}^{(1)}$ in terms of $R_\mathrm{ss}$, while the latter itself is given by $R_\mathrm{ss}=2 \pi~\mathrm{Re}[P_\mathrm{r,ss}^{(1)}]$, see Eq.~\eqref{eq:rss_BMF_lambda}. In this way, we obtain a self-consistent equation for $R_\mathrm{ss}$, which is to be solved numerically for given values of $D$ and $\lambda$.
The results thus obtained for $R_\mathrm{ss}$ are compared in Fig.~\ref{fig:BMF_globalresettting} against simulation results.  We observe a good agreement between theoretical and numerical estimates both for small and large values of $\lambda$. However, for intermediate $\lambda$, there is a noticeable deviation between the two, which will be discussed in Sec.~\ref{sec:MF-validation}. Note that the simulation results presented correspond to a very large system size $N = 10^{4}$, for which finite-size corrections are negligible, and hence, the aforesaid deviation may be attributed to the mean-field approximation not being able to capture analytically the behavior of the stationary-state order parameter.

\section{Application to the Kuramoto model}
\label{sec:Kuramoto}
\subsection{Dynamics in absence of resetting}
\label{sec:Kuramoto-no-reset}
We will now consider the noiseless version of Eq.~\eqref{eq:Noisy-Kuramoto-eom-1}, which models the Kuramoto model. Proceeding as in Sec.~\ref{sec:NK-no-reset}, the dimensionless evolution equation reads as 
\begin{align}
    \frac{d\theta_i}{dt}=\sigma \omega_i+\frac{1}{N}\sum_{j=1}^N \sin(\theta_j-\theta_i).
    \label{Eq:kuramoto-eom}
\end{align}
It was the genius of Kuramoto to demonstrate using an ingenious approach that for unimodal $g(\omega)$, the stationary-$R$ undergoes a continuous phase transition as a function of $K$, from a low-$K$ incoherent phase to a high-$K$ synchronized phase across a critical threshold~\cite{kuramoto1984chemical} 
\begin{align}
K_c=\frac{2}{\pi g(0)}.
\label{eq:Kc}
\end{align} 
Let us now turn to our method of analysis in Sec.~\ref{sec:general-method} and discuss how one may obtain the above result and even more. The large-$l$ behavior of $S^{(l)}$ and $\beta^{(l)}$ for the case $D=0$ is pursued in Appendix~\ref{subsec:app2}, where we show that $\beta^{(l)}=0~\forall~l$. Then, using Eq.~\eqref{eq:the_ansatz_general} together with Eq.~\eqref{eq:zero-Fourier-coefficient} and the obvious omission of the subscript $\mathrm{r}$ yield $P_{\mathrm{ss}}^{(1)}(\omega)= S/2\pi$, where $S=S_{\pm}=(-i \sigma \omega \pm \sqrt{4 b^2-\sigma^2\omega^2})/2 b$, see Appendix~\ref{subsec:app2}. It then follows that 
\begin{align}
\mathrm{Re}[S]
&= 
\begin{cases} 
   \frac{\pm \sqrt{R_\mathrm{ss}^2-\sigma^2\omega^2}}{R_\mathrm{ss}}&~~ |\omega| < R_\mathrm{ss}/\sigma, \\
    0 &~~|\omega| > R_\mathrm{ss}/\sigma,
\end{cases}
\label{eq:Re[s]}
\end{align}
where we have used the fact that $b=R_\mathrm{ss}/2$. Now, from Eq.~\eqref{eq:r-eqn-continuum-NK-ss}, we have
\begin{align}
    R_\mathrm{ss}&= 2 \pi \int d\omega~g(\omega) \mathrm{Re}[P_\mathrm{ss}^{(1)}(\omega)]. 
    \label{eq:rss_g(omega)Re[s]}
\end{align}
Using $P_{\mathrm{ss}}^{(1)}(\omega)= S/2\pi$ in Eq.~\eqref{eq:rss_g(omega)Re[s]}, we get
\begin{align}
    R_\mathrm{ss}
    &=\int  d\omega~g(\omega)~\mathrm{Re} [S],
    \label{eq:rss_integral1}
\end{align}
which upon using Eq.~\eqref{eq:Re[s]} yields
\begin{align}
    R_\mathrm{ss}
    &=\pm \int_{-R_{\mathrm{ss}}/\sigma}^{R_{\mathrm{ss}}/\sigma} d\omega~g(\omega)~\frac{\sqrt{R_\mathrm{ss}^2-\sigma^2\omega^2}}{R_\mathrm{ss}}. 
    \label{eq:rss_integral2}
\end{align}
Solving the above self-consistent equation yields $R_\mathrm{ss}$ for a given value of $\sigma$. Note that $R_\mathrm{ss}$ is always a solution of the equation.

To proceed, we will focus on two representative choices of frequency distribution for $g(\omega)$ (note that the analytical approach in Sec.~\ref{sec:general-method} is not limited to unimodal distributions, but applies equally well to bimodal distributions). The two choices are the Lorentzian distribution,
\begin{align}
    g(\omega)= \frac{1}{\pi}\frac{1/2}{(1/2)^2+\omega^2},\label{eq:Lorentzian}
\end{align}
and the uniform distribution,
\begin{align}
g(\omega)= \begin{cases}
       \frac{1}{2\sqrt{3}}~~\forall~|\omega| \leq \sqrt{3}, \\~
               0 ~~~\forall~~ |\omega| > \sqrt{3}.
         \end{cases} \label{eq:uniform_omega}
\end{align}
Below, we will discuss the form of $R_\mathrm{ss}$ for these two distributions by using Eq.~\eqref{eq:rss_integral2}.

\subsubsection{Case of Lorentzian $g(\omega)$:}
Here, we have the following equation determining $R_\mathrm{ss}$: 
\begin{align}
    R_\mathrm{ss}
    = \pm \frac{2}{\pi} \int_{-R_{\mathrm{ss}}/\sigma}^{R_{\mathrm{ss}}/\sigma} d\omega~\frac{\sqrt{R_\mathrm{ss}^2-\sigma^2\omega^2}}{R_\mathrm{ss}(1+4\omega^2)}.\label{eq:rss_integral3}
    \end{align}
Using $\omega=(R_{\mathrm{ss}}/\sigma)\sin\theta$ in Eq.~\eqref{eq:rss_integral3} finally yields
\begin{align}
    R_\mathrm{ss}
    &= \pm  \frac{2R_{\mathrm{ss}}}{\sigma-\sqrt{4R_\mathrm{ss}^2+\sigma^2}}.\label{eq:rss_integral4}
    \end{align}
For the negative sign on the RHS of Eq.~\eqref{eq:rss_integral4}, one gets 
\begin{align}
        &R_\mathrm{ss}^2(R_\mathrm{ss}^2-(1+\sigma))=0,
\end{align}
which, with $\sigma$ being an arbitrary positive quantity, does not always yield a non-zero solution of $R_\mathrm{ss}$ satisfying $0<R_\mathrm{ss}\le 1$. 
On the other hand, using the positive sign on the RHS of Eq.~\eqref{eq:rss_integral4}, one obtains  
\begin{align}
    &R_\mathrm{ss}^2(R_\mathrm{ss}^2-(1-\sigma))=0,
\end{align}
which, besides the zero-solution $R_\mathrm{ss}=0$, yields indeed a consistent solution of $R_\mathrm{ss}$, with $0<R_\mathrm{ss}\le 1$, as
\begin{align}
     R_\mathrm{ss}=\sqrt{1 - \sigma}.
     \label{eq:rnonzero-app-1}
\end{align}
Note that while the zero solution exists for all $\sigma$, the non-zero solution~\eqref{eq:rnonzero-app-1} exists so long as $\sigma\le \sigma_c=1$. In this case, $R_{\mathrm{ss}}$ undergoes a continuous transition from a synchronized to an incoherent phase, shown by the continuous line in Fig.~\ref{fig:Kuramoto_globalresettting}(a).

Recalling that the quantity $\sigma$ in Eq.~\eqref{eq:rnonzero-app-1} is the width of the frequency distribution scaled by the coupling constant $K$ (see the discussion preceding Eq.~\eqref{eq:NK-eom-1}), we obtain in terms of the unscaled width the synchronized solution for the stationary-$R$ in the case of the Lorentzian frequency distribution as 
\begin{align}
     R_\mathrm{ss}=\sqrt{1 - \sigma/K}.
     \label{eq:rnonzero-app}
\end{align}
The above result coincides with the one obtained within the OA ansatz~\cite{ott2008low} for the Kuramoto model. For a given value of $\sigma$, while the zero solution exists for all $K$, the non-zero solution~\eqref{eq:rnonzero-app} exists so long as $K\ge K_c=\sigma$. This critical value of $K$ is precisely the estimation of Kuramoto given in Eq.~\eqref{eq:Kc}. 

\subsubsection{Case of uniform $g(\omega)$:} For the uniform frequency distribution given by Eq.~\eqref{eq:uniform_omega}, the equation determining $R_\mathrm{ss}$ is obtained from Eq.~\eqref{eq:rss_integral2} as
\begin{align}
R_\mathrm{ss} &=\int_{-\sqrt{3}}^{\sqrt{3}} d\omega~\frac{1}{2\sqrt{3}}\sqrt{\frac{R_\mathrm{ss}^2-\sigma ^2 \omega^2}{R_\mathrm{ss}^2}}\nonumber\\
&=\frac{1}{2}\sqrt{1-\frac{3\sigma^2}{R_\mathrm{ss}^2}}+\frac{R_\mathrm{ss}} {2\sqrt{3}\sigma} \sin^{-1}{\frac{\sqrt{3}\sigma}{R_\mathrm{ss}}}\label{Eq:rss_uniform},
\end{align}
with the condition $R_\mathrm{ss} \ge \sqrt{3}\sigma$. For a given $\sigma$, the above self-consistent equation yields the value of $R_\mathrm{ss}$. A real solution exists provided the quantity under the square root is zero or positive, i.e., $\sigma \le R_\mathrm{ss}/\sqrt{3}$. Considering the upper limit on $\sigma$, we get the self-consistent solution as $R_\mathrm{ss}=\pi/4$, and, correspondingly, one has $\sigma=\pi/(4\sqrt{3})$. Thus, while the zero solution exists for all $\sigma$, a non-zero solution for $R_\mathrm{ss}$ exists so long as $\sigma \le \sigma_c=\pi/(4\sqrt{3})$. In contrast to the Lorentzian $g(\omega)$ considered above, the transition from the synchronized to the incoherent phase is now discontinuous, see Fig.~\ref{fig:Kuramoto_globalresettting}(b); the synchronized-phase $R_\mathrm{ss}$ values lie in the range $1\le R_\mathrm{ss} \le \pi/4$. We note that reverting to dimensional variables, Eq.~\eqref{Eq:rss_uniform} reads as 
\begin{align}
R_\mathrm{ss} &=\frac{1}{2}\sqrt{1-\frac{3\sigma^2}{K^2R_\mathrm{ss}^2}}+\frac{KR_\mathrm{ss}} {2\sqrt{3}\sigma} \sin^{-1}{\frac{\sqrt{3}\sigma}{KR_\mathrm{ss}}}\label{Eq:rss_uniform-1},
\end{align}
which matches with known results~\cite{PhysRevE.72.046211}. Also, reverting to dimensional variables, we obtain the critical coupling to obtain the synchronized phase as $K_c=4\sqrt{3}\sigma/\pi$. 

\subsection{Dynamics in presence of resetting}
\label{sec:Kuramoto-reset}
The Kuramoto model subject to stochastic resetting follows the dynamics~\eqref{eq:global_reset-protocol_noisyKuramoto} with $D=0$. The stationary-state density $P_\mathrm{r,ss}(\theta,\omega)$ satisfies the following equation (as obtained from Eq.~\eqref{eq:NK-reset-FP-ss} with $D=0$):
\begin{align}
   &\frac{\partial[(-\sigma \omega+R_\mathrm{ss}\sin \theta)P_\mathrm{r,ss}]}{\partial \theta}-\lambda P_\mathrm{r,ss}+\lambda \delta(\theta)=0
    \label{Eq:stationary_state_Kuramoto-reset},
\end{align}
where $R_\mathrm{ss}$ satisfies Eq.~\eqref{eq:rss_in_terms_of_P_{r,ss}^1}.

\begin{figure}[htbp!]
    \centering
\includegraphics[width=0.8\linewidth]{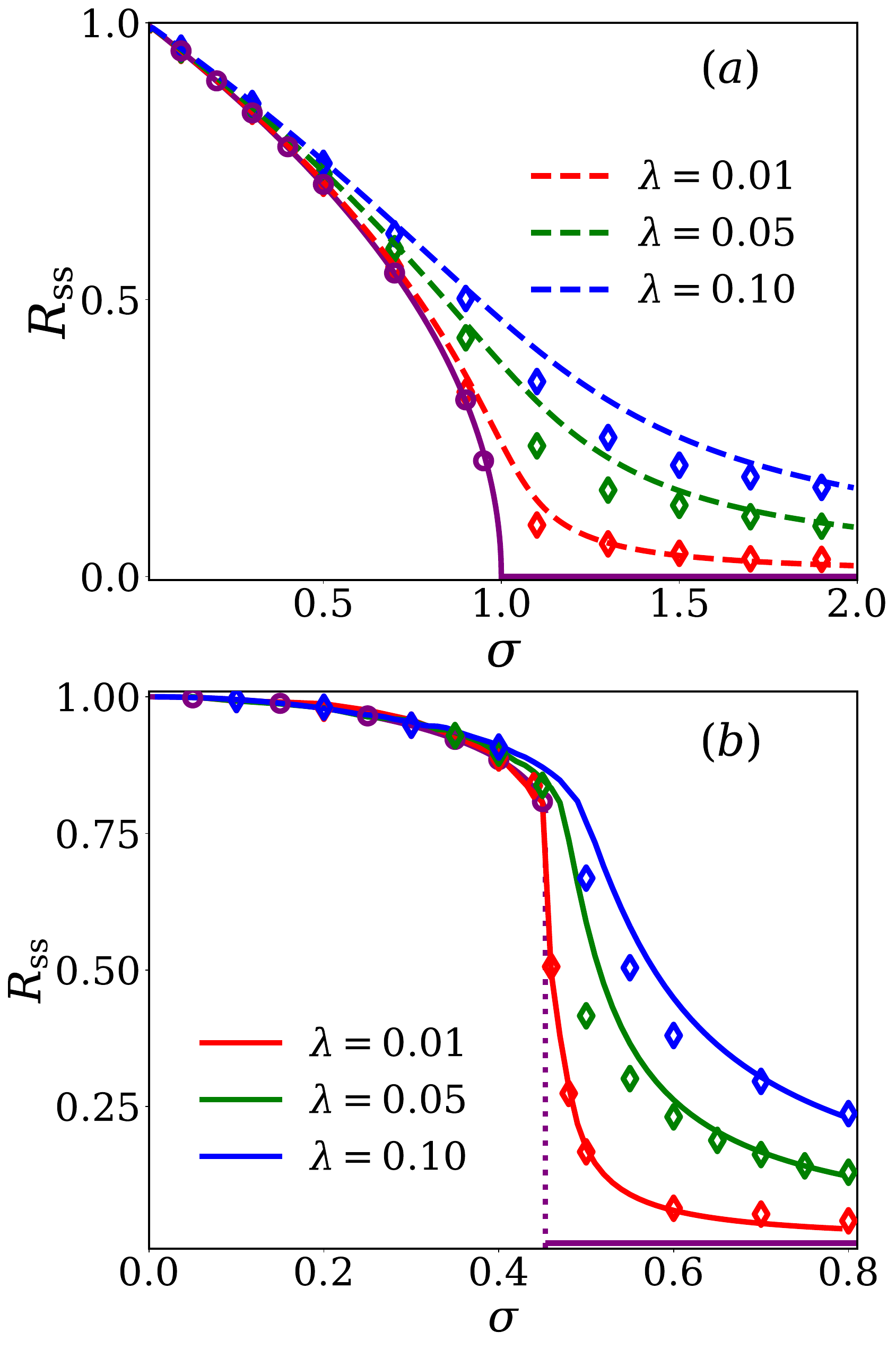}
    \caption{For the Kuramoto model discussed in Sec.~\ref{sec:Kuramoto}, panel (a) for the choice of the Lorentzian distribution~\eqref{eq:Lorentzian} for the natural frequencies shows the behavior of $R_\mathrm{ss}$ versus $\sigma$ in absence (continuous line) and presence (dashed line) of resetting. The former is given by Eq.~\eqref{eq:rnonzero-app-1}, while the latter is obtained on the basis of analysis in Section~\ref{sec:Kuramoto-reset}. The reset rates are: $\lambda = 0.01, 0.05, 0.10$. Panel (b) shows results, again in absence (continuous line) and presence (dashed line) of resetting, for the uniform distribution~\eqref{eq:uniform_omega} for the natural frequencies, and with the reset rate being $\lambda=0.01, 0.05, 0.10$. Here, the results in absence of resetting are obtained from Eq.~\eqref{Eq:rss_uniform}, while those in its presence are obtained on the basis of analysis in Section~\ref{sec:Kuramoto-reset}. The dotted line shows an abrupt jump in $R_\mathrm{ss}$ from the value $\pi/4$ to the value zero, implying a discontinuous transition. Simulation results are shown with unfilled diamonds for $\lambda \neq 0$ and unfilled circles for the bare model, both obtained using $N = 10^4$ oscillators; the data correspond to averaging over $50$ dynamical realizations.}
    \label{fig:Kuramoto_globalresettting}
\end{figure}

Proceeding as in Sec.~\ref{sec:BMF-reset}, we may solve Eq.~\eqref{Eq:stationary_state_Kuramoto-reset} by using the method detailed in Sec.~\ref{sec:general-method}, and by considering $D = 0$, $\lambda \ne 0$, and $\sigma \ne 0$. We show in Appendix~\ref{subsec:app3} that $S^{(l)}$ approaches a constant for large $l$, while $\beta^{(l)}$ vanishes as $l \to \infty$. By truncating the infinite series in Eq.~\eqref{eq:NK-Fourier}, now expressed for $P_\mathrm{r,ss}(\theta,\omega)$, and using Eq.~\eqref{eq:SNbetaN-main} with $D=0$, one obtains $P_\mathrm{r,ss}^{(1)}$ in terms of $R_\mathrm{ss}$. Equation~\eqref{eq:rss_in_terms_of_P_{r,ss}^1} then yields a self-consistent integral equation for $R_\mathrm{ss}$, which needs to be solved numerically for given values of $\sigma$ and $\lambda$. The results for $R_\mathrm{ss}$ are compared in Fig.~\ref{fig:Kuramoto_globalresettting} with numerical simulation results for the Lorentzian and the uniform $g(\omega)$ given respectively by Eqs.~\eqref{eq:Lorentzian} and \eqref{eq:uniform_omega}. As before, the agreement is quite good both for small and large values of $\lambda$, while a noticeable deviation is evident for intermediate $\lambda$. The deviation will be discussed in detail in Sec.~\ref{sec:MF-validation}. Here, we remark that this dynamical set-up, namely, the Kuramoto model under stochastic resetting, has been recently studied in Ref.~\cite{sarkar2022synchronization,bressloff2024global}. However, our present work focuses on the average $R_\mathrm{ss}$, and not on the full stationary distribution of the order parameter $R_\mathrm{ss}$ obtained across different realizations of the reset dynamics.

\section{Application to the Noisy Kuramoto model}
\label{sec:Noisy_Kuramoto}
\subsection{Dynamics in absence of resetting}
\label{sec:Noisy_Kuramoto-no-reset}
In this case, $P_\mathrm{ss}(\theta,\omega)$ is obtained as the solution of Eq.~\eqref{eq:NK-FP-ss}, given by~\cite{sakaguchi1988cooperative}
\begin{align}
&P_\mathrm{ss}(\theta,\omega)=e^{\frac{-R_\mathrm{ss}+ \sigma\omega\theta+R_\mathrm{ss}\cos \theta}{D}}P_\mathrm{ss}(0,\omega)\nonumber \\
&\times \left(1+\frac{(e^{-\frac{2 \pi \sigma\omega}{D}}-1)\int_0^\theta d\theta' ~e^{\frac{-\sigma\omega \theta'-R_\mathrm{ss} \cos \theta'}{D}}}{\int_{-\pi}^{\pi} d\theta'~e^{\frac{-\sigma\omega \theta'-R_\mathrm{ss}\cos\theta'}{D}}} \right),
\label{eq:ss-NK}
\end{align}
where $P_\mathrm{ss}(0,\omega)$ is a constant to be fixed from the normalization of $P_\mathrm{ss}(\theta,\omega)$. Note that on putting $\sigma$ to zero, the stationary state~\eqref{eq:ss-NK} reduces to the one given in Eq.~\eqref{eq:BMF-stationary-state} for the BMF model, Eq.~\eqref{eq:BMF-eom}, as it should. Substituting Eq.~\eqref{eq:ss-NK} in the first of the two equations in Eq.~\eqref{eq:r-eqn-continuum-NK-ss} with obvious modification in the latter yields a self-consistent equation for $R_\mathrm{ss}$. Upon analyzing, it follows that in the stationary state, $R_\mathrm{ss}$ shows a continuous phase transition at a fixed $\sigma$, from a low-$D$ synchronized/clustered phase to a high-$D$ incoherent/declustered phase (see the continuous line in Fig.~\ref{fig:NK_globalresettting} (a)) at the critical noise strength $D_c(\sigma)$ that satisfies the equation~\cite{sakaguchi1988cooperative}
\begin{align}
    2=\int_{-\infty}^\infty d\omega~\frac{D_c(\sigma)g(\omega)}{(D_c(\sigma))^2+\omega^2\sigma^2}.\label{eq:noisy_kuramoto_critical_D}
\end{align}
For the Lorentzian $g(\omega)$ in Eq.~\eqref{eq:Lorentzian},  the integral in Eq.~\eqref{eq:noisy_kuramoto_critical_D} can be evaluated by transforming it into a contour integral in the upper half of the complex $\omega$-plane and applying the residue theorem, resulting in
\begin{align}
    D_c= \frac{1}{2}-\frac{\sigma}{2}.\label{eq:NK_lorentzian_criticalD_noreset}
\end{align}
For the bare Kuramoto model, the above equation correctly reproduces the critical value $\sigma_c=1$, see the text following Eq.~\eqref{eq:rnonzero-app-1}. On the other hand, setting $\sigma$ to zero recovers correctly the critical point $D_c=1/2$ for the BMF model, see the text following Eq.~\eqref{eq:BMF-rss-consistent}.  
For uniform $g(\omega)$, as given by Eq.~\eqref{eq:uniform_omega}, Eq.~\eqref{eq:noisy_kuramoto_critical_D} yields 
$D_c$ as
\begin{align}
    D_c= \frac{\sqrt{3} \sigma}{\tan{2\sqrt{3} \sigma}}.\label{eq:NK_uniform_criticalD_noreset}
\end{align}
That $R_{\mathrm{ss}}$ undergoes a continuous transition from a synchronized to an incoherent phase in both of the above cases is shown in Fig.~\ref{fig:NK_globalresettting}.

In the present case, the aforementioned behavior of $R_\mathrm{ss}$ can  be obtained by our method discussed in Sec.~\ref{sec:general-method}, and by considering $D \ne 0$, $\sigma \ne 0$ and $\lambda = 0$. We can truncate the infinite series in Eq.~\eqref{eq:NK-Fourier}, now expressed for $P_\mathrm{ss}(\theta,\omega)$, and use Eq.~\eqref{eq:SNbetaN-main} with $\lambda=0$. This leads to $P_\mathrm{ss}^{(1)}$ being expressed in terms of $R_\mathrm{ss}$, which upon using in Eq.~\eqref{eq:r-eqn-continuum-NK-ss} yields a self-consistent integral equation for $R_\mathrm{ss}$. Solving this numerically for given $\sigma$, $D$ and $g(\omega)$ yields the aforementioned behaviour of $R_\mathrm{ss}$.

\subsection{Dynamics in presence of resetting}
\label{sec:noisy-Kuramoto-reset}
\begin{figure}[htbp!]
    \centering
\includegraphics[width=0.8\linewidth]{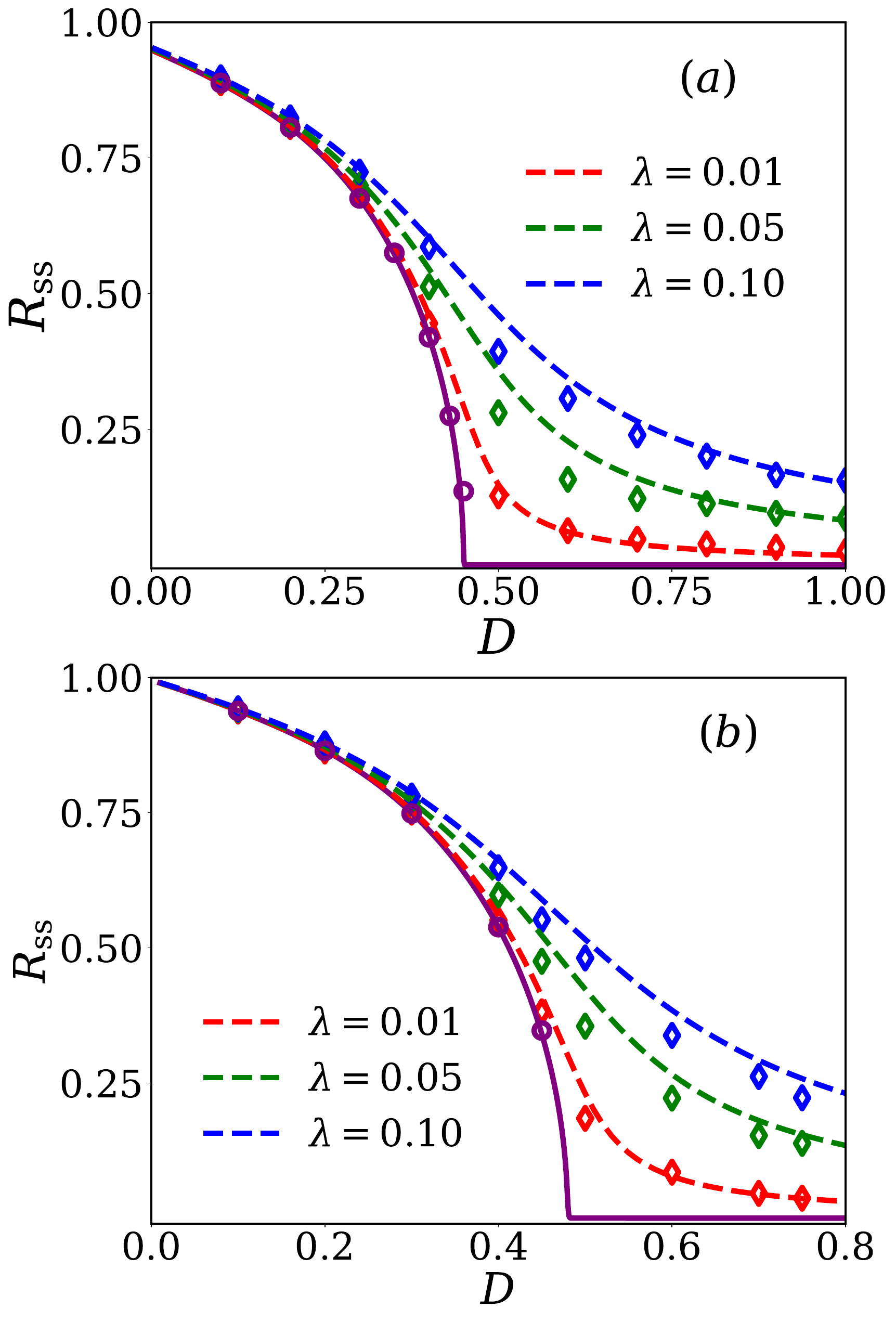}
    \caption{For the noisy Kuramoto model discussed in Sec.~\ref{sec:Noisy_Kuramoto}, panel (a) shows the behavior of $R_\mathrm{ss}$ versus $D$ in the absence (continuous line) and presence (dashed line) of resetting, with the natural frequencies of the oscillators chosen from the Lorentzian distribution~\eqref{eq:Lorentzian}. The results in absence of resetting are obtained on the basis of the analysis presented in Section~\ref{sec:Noisy_Kuramoto-no-reset}, while those in its presence are obtained on the basis of the analysis presented in Section~\ref{sec:noisy-Kuramoto-reset}. The parameters are $\lambda = 0.01,0.05,0.10$, $\sigma=0.1$. Panel (b) shows results for $R_\mathrm{ss}$ versus $D$ for the case of uniformly-distributed natural frequencies, \eqref{eq:uniform_omega}, with $\sigma=0.1$ and reset rates $\lambda=0.01,0.05,0.10$. As in panel (a), the results in absence of resetting are obtained on the basis of the analysis presented in Section~\ref{sec:Noisy_Kuramoto-no-reset}, while those in its presence are obtained on the basis of the analysis presented in Section~\ref{sec:noisy-Kuramoto-reset}. Simulation results in both the panels are shown with unfilled diamonds for $\lambda \neq 0$ and unfilled circles for the bare model, both obtained using $N = 10^4$ oscillators; the data correspond to averaging over $50$ dynamical realizations. }
    \label{fig:NK_globalresettting}
\end{figure}

This section discusses the dynamical set-up of the noisy Kuramoto model involving stochastic resetting, described by the dynamics~\eqref{eq:global_reset-protocol_noisyKuramoto}. Its stationary state density $P_\mathrm{r,ss}(\theta,\omega)$  follows Eq.~\eqref{eq:NK-reset-FP-ss}, and its Fourier-expansion coefficients satisfy Eq.~\eqref{eq:recurrence_global_reset_general}. Equation~\eqref{eq:NK-reset-FP-ss} may be solved using our method developed in Sec.~\ref{sec:general-method}, with $D \ne 0$, $\lambda \ne 0$, and $\sigma \ne 0$. As before, one has to first truncate the infinite series in Eq.~\eqref{eq:NK-Fourier} and express $P_\mathrm{r,ss}^{(1)}$ in terms of $R_\mathrm{ss}$. Equation~\eqref{eq:rss_in_terms_of_P_{r,ss}^1} then yields a self-consistent integral equation, which must be solved numerically to obtain $R_\mathrm{ss}$ for given values of $D$, $\sigma$ and $\lambda$. Figure~\ref{fig:NK_globalresettting} shows the results thus obtained for $R_\mathrm{ss}$ for the Lorentzian and the uniform $g(\omega)$ given by Eqs.~\eqref{eq:Lorentzian} and \eqref{eq:uniform_omega}, respectively. Similar to results obtained for the BMF and the Kuramoto model, a comparison between theoretical and numerical results shows good agreement at both small and large values of $\lambda$, but not so is the case for its intermediate values, an issue that will be discussed in Sec.~\ref{sec:MF-validation}.

\section{On the validity of the mean-field approximation}
\label{sec:MF-validation}
\begin{figure*}[htbp!]
\includegraphics[width=1\linewidth]{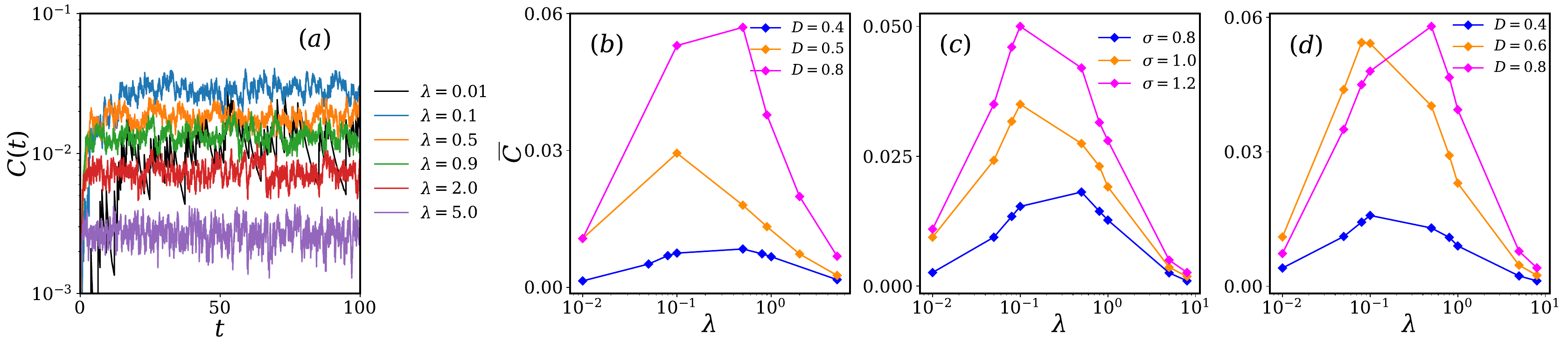}
    \caption{Figure~(a) shows the time evolution of the quantity $C(t)$ in Eq.~\eqref{eq:connected_correlation} for the BMF model in presence of resetting, with noise strength $D = 0.5$ and for different values of $\lambda$. The long-time average of $C(t)$ is shown as a function of $\lambda$ for the BMF model (panel~(b)), the Kuramoto model (panel~(c)), and the noisy Kuramoto model (panel~(d)). The results are obtained on the basis of numerical simulations using $N=10^4$ oscillators; the data correspond to averaging over $20$ dynamical realizations. For panels (c) and (d), we have used the Lorentzian distribution~\eqref{eq:Lorentzian} for the natural frequencies of the oscillators.}

    \label{fig:correlations_all_models}
\end{figure*}
This section explores the validity of the mean-field approximation used in our analysis. To this end, let us estimate the correlations that were neglected in implementing the mean-field approximation, by examining the quantity
\begin{align}
    &C(t) \equiv \langle |\mathbf{R}|^2\rangle(t)-|\langle \mathbf{R}\rangle(t)|^2, \label{eq:connected_correlation} 
\end{align}
with the terms on the RHS defined as
\begin{align}
    &\langle |\mathbf{R}|^2\rangle(t)\nonumber\\
    &\equiv \int d \omega' d\omega~d\theta~d\theta'~g(\omega) g(\omega')e^{i(\theta-\theta')}P_2((\theta,\omega),(\theta',\omega'),t)\nonumber \\
    &=\int\!\!d \omega' d\omega~d\theta~d\theta'~g(\omega) g(\omega')\cos(\theta-\theta')P_2((\theta,\omega),(\theta',\omega'),t), \label{eq:R2-defn} \\
    &\langle \mathbf{R}\rangle(t)\equiv \int d\omega~d\theta~g(\omega)e^{i\theta}P(\theta,\omega,t).
\end{align}
In obtaining the last equality in Eq.~\eqref{eq:R2-defn}, we have used the following results:
\begin{align}
    &\int\!\!d \omega' d\omega~d\theta~d\theta'~g(\omega) g(\omega')\sin(\theta-\theta')P_2((\theta,\omega),(\theta',\omega'),t) \nonumber \\
    &=\int\!\!d \omega' d\omega~d\theta~d\theta'~g(\omega) g(\omega')\sin(\theta'-\theta)P_2((\theta',\omega'),(\theta,\omega),t) \nonumber \\
    &=\int\!\!d \omega' d\omega~d\theta~d\theta'~g(\omega) g(\omega')\sin(\theta'-\theta)P_2((\theta,\omega),(\theta',\omega'),t),
\end{align}
implying that we have $\int~d \omega' d\omega~d\theta~d\theta'~g(\omega) g(\omega')\sin(\theta-\theta')P_2((\theta,\omega),(\theta',\omega'),t)=0$. Here, in obtaining the last step, we have used Eq.~\eqref{eq:P2-symmetry}. Note that, under the mean-field approximation, Eq.~\eqref{eq:P2} holds, and we have $C(t)=0$. When evaluating $C(t)$ in presence of resetting, we should replace $P_2((\theta,\omega),(\theta',\omega'),t)$ by $P_{2,\mathrm{r}}((\theta,\omega),(\theta',\omega'),t)$ and $P(\theta,\omega,t)$ by $P_\mathrm{r}(\theta,\omega,t)$. Needless to say, when evaluating correlations in the stationary state, we should use $P_{2,\mathrm{r,ss}}(\theta,\omega,t)$ in place of $P_{2,\mathrm{r}}(\theta,\omega,t)$, and so on.

Figure~\ref{fig:correlations_all_models} presents our numerical results on $C(t)$. Panel (a) shows for the BMF model that $C(t)$ for a fixed $D$ and for different resetting rates $\lambda$ approaches a stationary value at long times; this feature extends to the two other models of our study. Let us therefore focus on the stationary-state correlations, which we denote by 
\begin{align}
  \overline{C} \equiv \lim_{t \to \infty}C(t).  
\end{align}
While panel (b) displays for the BMF model the behavior of $\overline{C}$ as a function of $\lambda$, the same for the Kuramoto and the noisy Kuramoto model and for the Lorentzian $g(\omega)$ in Eq.~\eqref{eq:Lorentzian} are shown in panels (c) and (d). A key observation is that in all cases, the quantity $\overline{C}$ exhibits a non-monotonic dependence on $\lambda$, being relatively smaller at low and high $\lambda$ and larger at intermediate $\lambda$.

In order to understand how and why resetting induces correlations, we adapt the arguments presented in Ref.~\cite{biroli2023extreme} to the case at hand. We discuss the case of the BMF model but with $K=0$ (non-interacting particles), see Eq.~\eqref{eq:noisy-kuramoto-eom} in which one has to set $\omega_i=0~\forall~i$ and $K=0$. Our objective is to show that even in this non-interacting case in which there are no correlations otherwise, introducing resetting induces correlations in the stationary state, with qualitatively same non-monotonic dependence of the quantity $\overline{C}$ with respect to $\lambda$ as seen in Fig.~\ref{fig:correlations_all_models}, panels (b) -- (d). To this end, we note that the conditional probability density in presence of resetting, $P_{N,\mathrm{r}}(\{\theta_i\},t|\{\theta_i^{(0)}\},t'<t)$, may be obtained in terms of the conditional probability density in absence of resetting, $P_N(\{\theta_i\},t|\{\theta_i^{(0)}\},t'<t)$, using the same line of reasoning as the one invoked in deriving Eq.~\eqref{eq:reset_renewal_ring_1}. Here, $\{\theta_i^{(0)}\}$ stands for the initial condition. Thus, one has
\begin{align}
  &P_{N,\mathrm{r}}(\{\theta_i\},t|\{\theta_i^{(0)}\},0)=P_N(\{\theta_i\},t|\{\theta_i^{(0)}\},0)e^{-\lambda t}\nonumber \\
  &+\lambda\int_0^t d\tau~e^{-\lambda\tau}P_N(\{\theta_i\},t|\{\theta_i=\theta^\mathrm{r}\},t-\tau). 
  \label{eq:Ct-ss}
\end{align}
Here, we have used the fact that according to our resetting protocol, Eq.~\eqref{eq:global_reset-protocol_noisyKuramoto}, all particles undergo simultaneous resetting of their phase values to the common value $\theta^{\mathrm{r}}=0$.

\begin{figure}[htbp!]
\includegraphics[width=0.8\linewidth,height=0.6\linewidth]{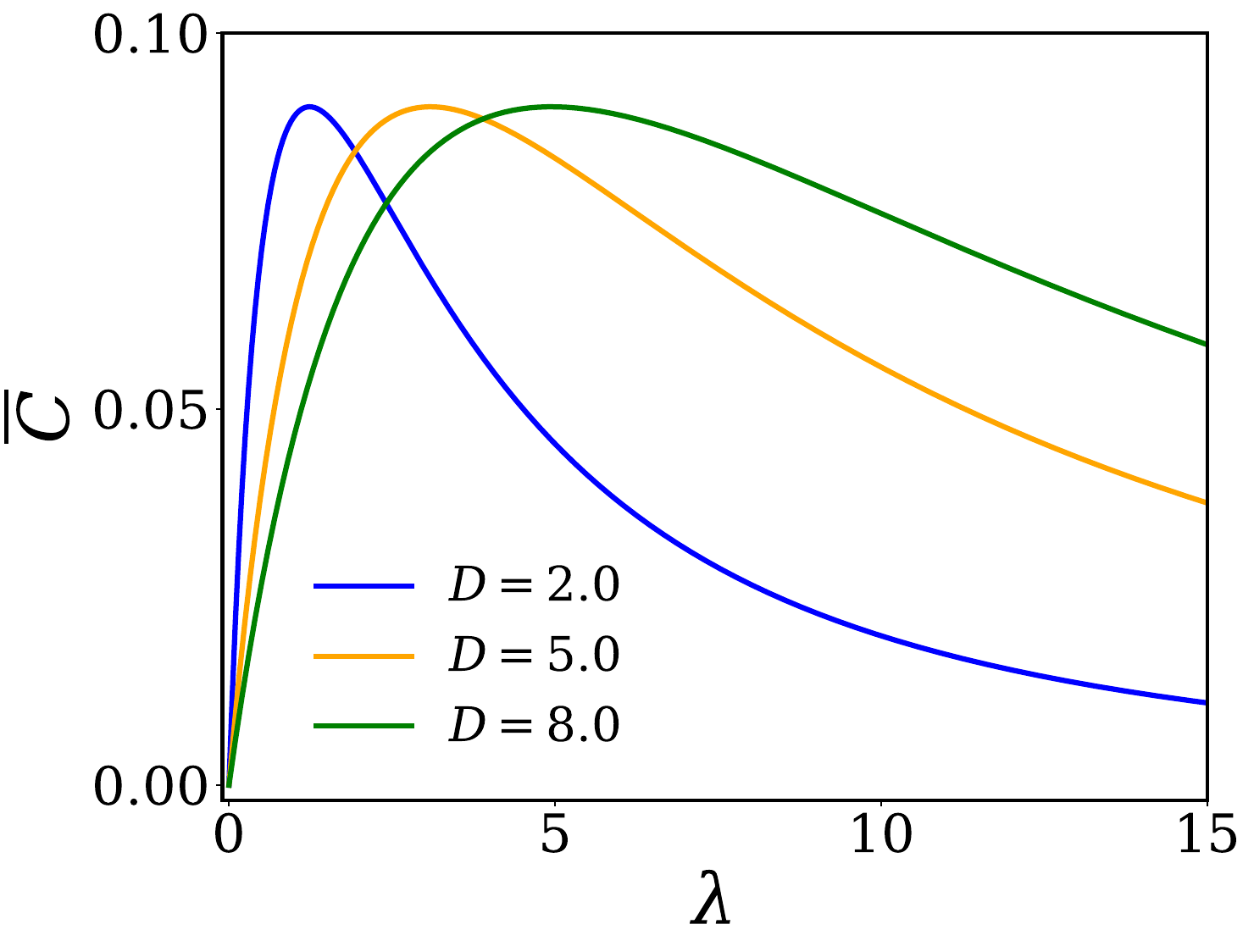}
    \caption{Plot of the function $\overline{C}$ in Eq.~\eqref{eq:Ct-final} with respect to resetting rate $\lambda$ for various values of the constant $D$.}
\label{fig:C-K0}
\end{figure}

In the stationary state, attained as $t\to \infty$, one obtains from Eq.~\eqref{eq:Ct-ss} that
\begin{align}
  &P_{N,\mathrm{r,ss}}(\{\theta_i\})=\lambda\lim_{t\to \infty}\int_0^t d\tau~e^{-\lambda\tau}P_N(\{\theta_i\},t|\{\theta_i=\theta^\mathrm{r}\},t-\tau).  
\end{align}
Since the particles are non-interacting, we may use the results of Section~\ref{sec:single-particle-no-reset}. In particular, using Eq.~\eqref{eq:free_prop_ring}, we get
\begin{align}
  &P_{N,\mathrm{r,ss}}(\{\theta_i\})=\lambda\left(\frac{1}{2 \pi}\right)^N\sum_{l_1,l_2,\ldots,l_N=-\infty}^\infty e^{i (l_1\theta_1+\ldots l_N\theta_N)}\nonumber \\
  &\times\int_0^\infty d\tau~e^{-(\lambda+D(l_1^2+\ldots+l_N^2))\tau}\nonumber \\
  &=\lambda\left(\frac{1}{2 \pi}\right)^N\sum_{l_1,l_2,\ldots,l_N=-\infty}^\infty \frac{e^{i (l_1\theta_1+\ldots l_N\theta_N)}}{\lambda+D(l_1^2+\ldots+l_N^2)}.
\end{align}
It then follows that $P_{N,\mathrm{r,ss}}(\{\theta_i\})$ does not factorize into a product of single-particle probability densities, implying correlations between the particles induced solely by the act of resetting. We get
\begin{align}
    P_{2,\mathrm{r,ss}}(\theta_1,\theta_2)&=\int_{-\pi}^\pi \ldots \int_{-\pi}^\pi d\theta_3\ldots d\theta_N~P_{N,\mathrm{r,ss}}(\{\theta_i\})\nonumber \\
  &=\lambda\left(\frac{1}{2 \pi}\right)^2\sum_{l_1,l_2=-\infty}^\infty \frac{e^{i (l_1\theta_1+ l_2\theta_2)}}{\lambda+D(l_1^2+l_2^2)}, \\
  P_\mathrm{r,ss}(\theta_1)&=\int_{-\pi}^\pi \ldots \int_{-\pi}^\pi d\theta_2\ldots d\theta_N~P_{N,\mathrm{r,ss}}(\{\theta_i\})\nonumber \\
  &=\lambda\left(\frac{1}{2 \pi}\right)\sum_{l=-\infty}^\infty \frac{e^{i l\theta_1}}{\lambda+Dl^2}.
\end{align}
The last equation above matches with Eq.~\eqref{reset_renewal_ring-ss}, as it should. From the form of $P_{N,\mathrm{r,ss}}(\theta_1,\theta_2)$ given above, it is evident that the factorization of the form in Eq.~\eqref{eq:factorization} does not hold: $P_{2,\mathrm{r,ss}}(\theta_1,\theta_2)\ne P_\mathrm{r,ss}(\theta_1)P_\mathrm{r,ss}(\theta_2)$.
We then get
\begin{align}
    &\int_{-\pi}^\pi \int_{-\pi}^\pi d\theta_1 d\theta_2~\cos(\theta_1-\theta_2)P_{2,\mathrm{r,ss}}(\theta_1,\theta_2)=\frac{\lambda}{\lambda+2D}, \\
    &\int_{-\pi}^\pi d\theta~e^{i\theta}P_\mathrm{r,ss}(\theta)=\frac{\lambda}{\lambda+D},
\end{align}
yielding
\begin{align}
    \overline{C}=\frac{\lambda}{\lambda+2D}-\left(\frac{\lambda}{\lambda+D}\right)^2.
    \label{eq:Ct-final}
\end{align}
Plotting $\overline{C}$ with respect to $\lambda$, one may observe from Fig.~\ref{fig:C-K0} qualitatively similar non-monotonic dependence with respect to $\lambda$ as seen in Fig.~\ref{fig:correlations_all_models}, panels (b) -- (d). In the absence of any explicit results for the conditional probability density $P_{N,\mathrm{r}}(\{\theta_i\},t|\{\theta_i^{(0)}\},t'<t)$ in presence of interactions ($K \ne 0$), the analysis above cannot be pursued for the case of the interacting particle systems considered in this work. From Fig.~\ref{fig:C-K0}, since correlations are most pronounced at intermediate values of the resetting rate $\lambda$, where the very tenet of our analytical calculation, Eq.~\eqref{eq:factorization}, does not hold, we now understand at least phenomenologically the mismatch between theory and simulation results observed in Figs.~\ref{fig:BMF_globalresettting}, \ref{fig:Kuramoto_globalresettting}, \ref{fig:NK_globalresettting}. Note that despite the mismatch, the theory is able to predict correctly the qualitative behavior of the stationary-state order parameter across the models of our study.

\section{Conclusions}
\label{sec:Conclusion}

In this work, we investigated the role of global resetting in a variety of dynamical systems, ranging from a single particle on a circle to many-body interacting systems modeled by Kuramoto-type dynamics involving coupled limit-cycle oscillators evolving both in absence and presence of noise. Using a mean-field approximation, we developed an analytical technique to determine across the studied set-ups the stationary-state order parameter. We reported that the mean-field approximation works perfectly well in absence of resetting, but provides qualitative match with simulation results in presence of resetting as a consequence of strong correlations induced by resetting due to which the approximation fails at intermediate resetting rate values. Our study provides insights into how an interplay between the intrinsic dynamics and resetting influences collective synchronization in models of limit-cycle oscillators within the framework of the Kuramoto model. Here, we considered global resetting, whereby all the constituent degrees of freedom of the system undergo simultaneous resetting to a common reset state. It would be interesting to consider suitable refinement of the mean-field approximation invoked in this work in order to have a better match of analytical with simulation results. An immediate extension is to consider resetting only a subset of the degrees of freedom at random times~\cite{MajumderSubsystemReset2024, zhao2024subsystem, bressloff2024kuramoto}, a set-up whose analysis using the analytical framework developed here will be reported elsewhere. Other future directions include extending our formalism to spatially-extended systems defined on a lattice, to the case of heterogeneous coupling between the degrees of freedom, and to non-Markovian resetting protocols.

\section{Acknowledgements}
M.S. acknowledges support by the Deutsche Forschungsgemeinschaft (DFG, German Research Foundation) under Germany’s Excellence Strategy EXC 2181/1-390900948 (the Heidelberg STRUCTURES Excellence Cluster). S.G. thanks ICTP–Abdus Salam International Centre for Theoretical Physics, Trieste, Italy, for
support under its Regular Associateship scheme, and for hospitality during March 2025 when the paper was finalized. We gratefully acknowledge the generous allocation of computing resources by the
Department of Theoretical Physics (DTP) of the Tata Institute of Fundamental Research (TIFR), and related technical
assistance from Kapil Ghadiali and Ajay Salve. This work is supported by the Department of Atomic Energy, Government of
India, under Project Identification Number RTI 4002.

\appendix
\section{Large-\texorpdfstring{$l$}{} behaviour of \texorpdfstring{$S^{(l)}$}{} and \texorpdfstring{$\beta^{(l)}$}{} with \texorpdfstring{$b\neq 0$}{}}

\label{sec1:app1}

We start with Eq.~\eqref{eq:continued_fraction_general} that gives the following expressions for $S^{(l)}$ and $\beta^{(l)}$: 
\begin{align}
    &S^{(l)}=\frac{b~l}{b~l S^{(l+1)} +(D l^2+i \sigma\omega l+\lambda)},\nonumber \\
    \label{eq:continued_fraction_general-app}\\
    &\beta^{(l)}=\frac{c- b~l \beta^{(l+1)}}{b~l S^{(l+1)} +(D l^2+i \sigma\omega l
    +\lambda)}. \nonumber
\end{align}
 Considering $b\ne 0$, we now want to argue that with increasing $l$, the quantities $S^{(l)}$ and $\beta^{(l)}$ converge to either zero or a non-zero constant. Assume that convergence happens for a large value of $l$, say $l=L$, such that $S^{(L)}=S^{(L+1)}$ and $\beta^{(L)}=\beta^{(L+1)}$. From Eq.~\eqref{eq:continued_fraction_general-app}, we get 
 \begin{align}
     &S^{(L)}=\frac{b~L}{b~L S^{(L)} +(D L^2+i \sigma\omega L+\lambda)},\nonumber \\
     \label{eq:continued_fraction_general-1-app}
     \\
    &\beta^{(L)}=\frac{c- b ~ L \beta^{(L)}}{b~L S^{(L)} +(D L^2+i \sigma\omega L
    +\lambda)}. \nonumber
 \end{align}
 We will now consider $3$ cases of relevance in the following.
 
\subsection{The case of \texorpdfstring{$D$, $\sigma$, $\lambda$}{} all non-zero}
\label{subsec:app1}
Here, Eq.~\eqref{eq:continued_fraction_general-1-app} gives
\begin{align}
   S^{(L)}= \frac{- H_{L} \pm \sqrt{H_{L}^2+4} }{2}=S^{(L)}_{\pm},
   \label{eq:app-SL}
\end{align}
where we have defined $H_L \equiv (D L^2+i \sigma\omega L+\lambda)/b L$.
We must decide now as to which one of the two roots  $S^{(L)}_{\pm}$ we should consider for further analysis.
As $L\to \infty$, we have $S^{(L)}_{-} \to - D L$, thus contradicting our convergence assumption of $S^{(L)}=S^{(L+1)}$. Hence, discarding $S^{(L)}_{-}$ and considering $S^{(L)}_{+}$, we have for large $L$ that
\begin{align}
    S^{(L)}
    &\approx\frac{- H_{L} + H_{L}(1+2\mathcal{O}(1/H_{L}^2))}{2},
    \label{eq:S_large_l_behaviour}
\end{align}
so that $H_{L}^{-1} \to 0$ as $L \to \infty$ implies  $S^{(L\to \infty)}=0$. 

Let us discuss the large-$l$ behaviour of $\beta^{(l)}$. We have from Eq.~\eqref{eq:continued_fraction_general-1-app} that
\begin{align}
    \beta^{(L)}&= \frac{c}{b~L (S^{(L)}+1)+(D L^2+i \sigma\omega L
    +\lambda)}.
    \label{eq:betaL-app}
\end{align}
As $L \to \infty$, when $S^{(L)} \to 0$, the presence of the term $D L^2$ in the denominator on the RHS implies $\beta^{(L)} \to 0$. We thus conclude that so long as $D$, $\sigma$, $\lambda$ are all non-zero, both $S^{(l)}$ and $\beta^{(l)}$ for a given $\omega$ decrease with increasing $l$, decaying to zero for large $l$. Redoing the aforesaid analysis, one finds that it holds true so long as $D\ne 0$. 

\subsection{The case of \texorpdfstring{$D=0$, $\lambda=0$, $\sigma \neq 0$}{}}
\label{subsec:app2}
In this case, Eq.~\eqref{eq:continued_fraction_general-app} yields that $S^{(l)}$ becomes $l$-independent, and is given by 
\begin{align}
    S=S_{\pm}=\frac{-i \sigma \omega \pm \sqrt{4 b^2-\sigma^2\omega^2}}{2 b}.
    \label{eq:SL-D0lamb0}
\end{align}
The appropriate choice of $S$ among the two possible roots $S_{\pm}$ has been discussed in detail in Sec.~\ref{sec:Kuramoto-no-reset}. Equation~\eqref{eq:continued_fraction_general-app} with $c=\lambda/(2\pi)=0$ makes also $\beta^{(l)}$ an $l$-independent quantity, and which satisfies 
\begin{align}
    (b~ S+i \sigma\omega)\beta+b ~ \beta=0,\label{eq:app_be_zero_condition}
\end{align}
implying thereby that $\beta=0$. 

\subsection{The case of \texorpdfstring{$D=0$, $\lambda \neq 0$, $\sigma \neq 0$}{D=0, lambda≠0, sigma ≠ 0}} 
\label{subsec:app3}
In this case, Eq.~\eqref{eq:app-SL} holds with 
\begin{align}
H_L=\frac{i \sigma\omega L+\lambda}{b L},   
\end{align}
so that $H_{L} \to i\sigma\omega/b$ as $L \to \infty$, implying for large $L$ that
\begin{align}
 S_\pm^{(L)}= \frac{-i\sigma \omega/b\pm \sqrt{4-\sigma^2\omega^2/b^2} }{2}.
 \end{align}
 Since $S^{(l)}$ approaches a constant for large $l$, using Eq.~\eqref{eq:betaL-app} with $D=0$ leads us to conclude that $\beta^{(l)}$ vanishes as $l \to \infty$.

\section{Resetting to \texorpdfstring{$0 \leq R^r < 1$}{R\^r less than 1}}
\label{sec2:app1}

In the main text, we have considered the case of global resetting of the oscillator phases to the common value $\theta^\mathrm{r}=0$, so that the value of the order parameter at reset is $R^\mathrm{r}=1$. We now discuss the case of resetting when the order parameter at reset takes up a value $0 \le R^\mathrm{r} <1$. One possibility is to reset a given fraction $\gamma$ of the oscillator phases to the value $\theta_\mathrm{r}=0$, while the remaining $(1-\gamma)$ fraction of oscillators are reset to the value $\theta_\mathrm{r}=\pi$, where $\gamma$ is related to $R^\mathrm{r}$ by 
\begin{align}
    R^\mathrm{r}= |2\gamma-1|.
\end{align}
In this case, Eq.~\eqref{eq:NK-reset-FP-ss} reads as
\begin{align}
    &-\frac{\partial[(\sigma \omega-R_\mathrm{ss} \sin \theta)P_\mathrm{r,ss}(\theta,\omega)]}{\partial \theta}+D\frac{\partial^2 P_\mathrm{r,ss}(\theta,\omega)}{\partial \theta^2}\nonumber \\
    &-\lambda P_\mathrm{r,ss}(\theta,\omega)+\lambda [\gamma \delta(\theta)+(1-\gamma)\delta(\pi-\theta)]=0.
    \label{eq:NK-reset-FP-ss_general}
\end{align}
Using Eq.~\eqref{eq:NK-Fourier}, we obtain the modified version of the recurrence relation Eq.~\eqref{eq:recurrence_global_reset_general} as
\begin{align}
&b~l P_\mathrm{r,ss}^{(l+1)}+(D l^2+i \sigma \omega l+\lambda) P_\mathrm{r,ss}^{(l)} - b~l P_{\mathrm{r,ss}}^{(l-1)}-c=0,
\label{eq:recurrence_global_reset_general_arbitrary_r0}\end{align}
where we have $b\equiv R_\mathrm{ss}/2$, as before, see Eq.~\eqref{eq:bc-defn}, but with the quantity $c$ now being defined as  
\begin{align}
 c\equiv\frac{\lambda}{2 \pi}( \gamma+(-1)^l (1-\gamma)).
\end{align}
Following the steps in Section~\ref{sec:general-method}, we obtain Eq.~\eqref{eq:Prss-Glbetal}, 
with the same $G^{(l)}$, $S^{(l)}$, $\beta^{(l)}$. One may check on putting $\gamma=1$ that one recovers the expressions given in Section~\ref{sec:general-method}. The corresponding stationary state order parameter is thus obtained as
\begin{align}
R_\mathrm{ss}&=2 \pi\int  d\omega~g(\omega)~\mathrm{Re}[P_\mathrm{r,ss}^{(1)}(\omega)].\label{eq:rss_in_terms_of_P_{r,ss}^1_arbitrary_r0}
\end{align}

\bibliography{revised_manuscript}

\end{document}